\documentclass[conference,compsoc]{IEEEtran}
\IEEEoverridecommandlockouts
\usepackage{xcolor,colortbl} 
\usepackage[linesnumbered,ruled,vlined]{algorithm2e}%
\usepackage{setspace}%
\usepackage{textcomp}

\newcolumntype{?}{!{\vrule width 1pt}}
\newcolumntype{P}[1]{>{\centering\arraybackslash}p{#1}}

\newcolumntype{L}[1]{>{\centering\arraybackslash}p{#1}}

\newcommand{\graycell}[0]{\cellcolor [gray]{0.85}}
\usepackage{url}

\usepackage{hyperref}
\usepackage{float}
\usepackage{amsmath,amssymb,amsfonts} %
\usepackage{graphicx}
\usepackage{multirow}%
\usepackage{microtype}
\usepackage{cite}
\usepackage{xurl}
\usepackage{bm}

\newcommand{\finetunned}[1]{\textcolor{blue}{#1}}

\newcommand{\revision}[1]{\textcolor{black}{#1}}
\newcommand{\toremove}[1]{\textcolor{green}{}}

\newcommand{\remove}[1]{\textcolor{red}{}}

\def\BibTeX{{\rm B\kern-.05em{\sc i\kern-.025em b}\kern-.08em
    T\kern-.1667em\lower.7ex\hbox{E}\kern-.125emX}}
\begin{document}

\title{A Fly on the Wall - Exploiting Acoustic Side-Channels in Differential Pressure Sensors
}

\author{\IEEEauthorblockN{Yonatan~Gizachew~Achamyeleh$^1$, Mohamad Habib Fakih$^1$, Gabriel Garcia$^1$, Anomadarshi Barua$^2$, \\ Mohammad Abdullah Al Faruque$^1$}
\IEEEauthorblockA{
{\textit{$^1$Dept. of Electrical Engineering and Computer Science, University of California, Irvine, CA, USA}} \\
{\textit{\{yachamye, mhfakih, gegarci1, alfaruqu\}@uci.edu}} \\
{\textit{$^2$Dept. of Cyber Security Engineering, George Mason University, VA, USA.}} 
{\textit{abarua8@gmu.edu}} 
}

}


\maketitle
\thispagestyle{plain}
\pagestyle{plain}
\begin{abstract}
Differential Pressure Sensors are widely deployed to monitor critical environments. However, our research unveils a previously overlooked vulnerability: their high sensitivity to pressure variations makes them susceptible to acoustic side-channel attacks. We demonstrate that the pressure-sensing diaphragms in DPS can inadvertently capture subtle air vibrations caused by speech, which propagate through the sensor's components and affect the pressure readings.
Exploiting this discovery, we introduce \textbf{BaroVox}, a novel attack that reconstructs speech from DPS readings, effectively turning DPS into ``a fly on the wall." We model the effect of sound on DPS, exploring the limits and challenges of acoustic leakage. To overcome these challenges, we propose two solutions: a signal-processing approach using a unique spectral subtraction method and a deep learning-based approach for keyword classification.
Evaluations under various conditions demonstrate BaroVox's effectiveness, achieving a word error rate of 0.29 for manual recognition and 90.51\% accuracy for automatic recognition. Our findings highlight the significant privacy implications of this vulnerability. 
We also discuss potential defense strategies to mitigate the risks posed by BaroVox.

\end{abstract}


\section{Introduction}
Differential Pressure Sensors (DPS) have become ubiquitous in various environments, ranging from industrial facilities and cleanrooms to residential buildings, offices, hotels, and hospitals~\cite{avnet2024pressure, standard2004cleanrooms, datasheetsdp}. These sensors are designed to measure minute pressure differences between two points, enabling precise control and monitoring of critical systems such as HVAC, airflow management, and room pressure regulation~\cite{bradford_hvac_2024, standard2004cleanrooms}. 
\revision{While DPS find applications in various sectors, their role is particularly critical in the semiconductor industry, where they are essential for maintaining cleanroom integrity.}
However, the widespread deployment of DPS has inadvertently introduced a hidden vulnerability that can be exploited for eavesdropping.

In many real-world applications, audio systems, including speakers and intercoms, are often installed in close proximity to DPS. This practice is driven by various factors, such as the need for effective communication, audio notifications, or entertainment purposes. For instance, intercoms are used in industrial cleanrooms to facilitate coordination among workers without compromising the controlled environment~\cite{standard2004cleanrooms}. 

While the co-location of audio systems and DPS serves practical purposes, it unintentionally creates an acoustic side channel that attackers can exploit. DPS's high sensitivity to pressure variations makes them susceptible to unintended acoustic coupling. When sound waves from nearby speakers or intercoms impinge upon the DPS, they induce minute vibrations on the sensor's diaphragm, causing measurable changes in its output. This unintended interaction effectively transforms the DPS into a makeshift microphone, allowing potential attackers to eavesdrop on confidential conversations or recover sensitive audio information.

In this paper, we introduce BaroVox, a novel side-channel attack that exploits DPS's acoustic vulnerability to recover speech from their output signals. BaroVox leverages audio systems' close proximity to DPS, which is a common deployment scenario in various real-world settings. By carefully analyzing the pressure variations captured by the DPS, BaroVox enables the reconstruction of speech signals, effectively turning these ubiquitous sensors into unintended listening devices.

\color{black}
We argue that if an attacker manages to get the pressure reading of this DPS, they can process the pressure data and partially reconstruct speech to still confidential information. 
The main challenge in realizing BaroVox lies in extracting intelligible audio from the low-bandwidth, noisy signals captured by DPS, which are primarily designed for measuring pressure differences rather than recording sound.
The sensor's non-linear frequency response adds another layer of complexity. 
BaroVox offers a unique \textbf{Pressure-Acoustic Transformation (PAT)} to mitigate these challenges, presenting two approaches for eavesdropping conversations.

The first design solution employs PAT for speech reconstruction and then focuses on enhancing the reconstructed speech's signal-to-noise ratio (SNR).
This enhancement is achieved by integrating multiple digital signal processing techniques, comprising a novel spectral subtraction method, normalization, high-pass filtering, and equalization.
In the spectral subtraction phase, first,
our design divides the reconstructed speech into percussive and harmonic segments using median filtering~\cite{fitzgerald2010harmonic, driedger2014extending}.
Then, after evaluating the statistical property of noise in the target environment, we apply tailored spectral noise removal to both components, adjusting the degree of removal for each before their reintegration. 
{This results in an improved SNR of the speech while keeping the integrity of other speech elements.}

The second design solution leverages deep learning techniques to extract pertinent audio features, classifying words from a specific vocabulary dataset. Our Automatic Speech Recognition (ASR) model introduces \textbf{denoising autoencoder} and \textbf{equalization
layers} on top of ResNet to address the challenges posed by low SNR and non-linear frequency response of the sensor. This solution can be preferred by attackers when more precision is desired, and focus is required on specific critical keywords. 

We extensively evaluate the effectiveness of BaroVox through real-world experiments, considering various factors that influence the success of the attack.
We evaluate the first solution's capability in Manual Speech Recognition (MSR) by engaging {18 volunteers to transcribe 20} reconstructed speeches from pressure data. 
The performance metrics are Mean Opinion Score (MOS)~\cite{leng2021mbnet} and Word Error Rate (WER)~\cite{errattahi2018automatic}. 
Our results demonstrate BaroVox's alarming efficacy. The proposed signal processing pipeline achieves a WER of 0.29, comprehending over 70\% of the discourse.
The second solution's efficacy was gauged on its word classification prowess within a restricted vocabulary dataset. 
Our ASR model achieved an impressive {{90.51}}\% accuracy for a 35-keyword, speaker-independent classification.

These findings highlight the serious privacy implications of BaroVox.
Attackers can exploit this vulnerability to eavesdrop on sensitive conversations, compromise confidential information, or invade privacy without physical access to the targeted premises.
The implications are particularly severe in critical environments such as industrial facilities, corporate offices, and healthcare institutions, where the loss of sensitive data can have far-reaching consequences.
Our main \textbf{contributions} are summarized as follows:

 \textbf{(1)} We discover a previously overlooked acoustic side-channel vulnerability in deploying DPS in proximity to sound sources.
 
\textbf{(2)} We characterize the limits and challenges of speech recovery from pressure reading. We propose \textbf{BaroVox} to overcome these challenges. To the best of our knowledge, \textit{BaroVox is the first side-channel attack on pressure sensors}.


\textbf{(3)} We evaluate BaroVox on MSR and ASR tasks and show an attacker can partially reconstruct a speech with $0.29$ WER using MSR and {$90.51\%$ accuracy} using ASR. 
We present sample reconstructed audio for demonstration here: {\href{https://sites.google.com/view/barovox-a-fly-on-the-wall/home}{\textcolor{blue}{{BaroVox}}}.

\textbf{(4)} 
We propose practical countermeasures and defense strategies to mitigate the risks associated with BaroVox.

\color{black}

\section{Background}

In this section, we discuss DPSs and their applications across various domains, as well as the privacy implications of deploying DPS in close proximity to audio systems.

\subsection{Physics of Differential Pressure Sensors}
\label{subsection:physicsofDPS}
The structure of a DPS is illustrated in Fig.~\ref{fig:thermalDPSComp}. A DPS consists of three fundamental components: 1) pressure ports, 2) a pressure-sensing element, and 3) a transducer. Pressure ports are openings in the sensor where pressure gets applied and are connected directly to the pressure-sensing element.

The pressure-sensing element is a force collector constructed of a flexible diaphragm such as a thin semiconductor material film, a silicon membrane, or another material that responds to the measured pressure. \textit{In our attack, we show that the thin and flexible nature of the diaphragm enables the sensing element to respond to tiny vibrations created by sound waves. We then utilize these vibrations collected by the sensing element to reconstruct speech.}

\remove{The transducer component transforms the pressure-sensing element's mechanical strain into a measurable and interpretable electrical signal.}
Various physical mechanisms, such as thermal mass-flow, capacitive sensing, or piezoresistive sensing, could form the basis of the transducer element.
Of all DPSs, thermal mass-flow and piezoresistive sensing-based DPSs are widely used in various applications. However, thermal mass-flow-based DPSs are preferable due to their high measuring accuracy even on long connecting hoses~\cite{firstsensor:2020}. For this reason, we used a thermal mass-flow DPS to show the feasibility of our attack.
Specifically, we utilized the SDP800~\cite{datasheetsdp} DPS from Sensirion, and we will provide a detailed explanation of its structure below.
\revision{\textit{Importantly, the vulnerability we have discovered is inherent to the fundamental design of DPS, specifically the flexible diaphragm used as a force collector. This diaphragm-based sensing mechanism is common across various DPS types. This makes all DPSs vulnerable to our attack as they can pick up tiny vibrations.}}

\textbf{Thermal mass-flow DPS}: Fig.~\ref{fig:thermalDPSComp} depicts the structure of a thermal mass-flow DPS. This DPS uses a pressure-sensing element which consists of a thin semiconductor diaphragm (Silicon Nitrate film), two temperature sensors, and a heating element.
As air moves across this diaphragm, it prompts a temperature differential between the temperature sensors.
This temperature shift, which correlates with the mass flow rate of the air, is then adeptly translated by the transducer into an electrical signal, reflecting the differential pressure.

\textbf{Electronics inside of DPSs:} DPS has other electronic circuits to condition and relay signals. These circuits process the signal derived from the sensing element, ultimately generating an output indicative of the measured differential pressure.
The circuitry amplifies the output signal, including any pressure differential captured by the pressure-sensing element due to sound vibration.
As illustrated in Fig.\ref{fig:thermalDPSComp}, this processing sequence involves components like an amplifier, a Digital Signal Processor (DSP), and an Analog-to-Digital Converter (ADC).
Once digitized, the signal is directed to a microcontroller via the Communication Interface (CI). The ADC's sampling rate might constrain data transfer to the controller, influencing the sampling rate of the pressure reading --- an aspect detailed in Sec.~\ref{subsection:challenges}.
\vspace{-.50em}
\begin{figure}[h!]
\centering
\includegraphics[trim=1pt 1pt 1pt 25pt,clip,width=0.50\textwidth]{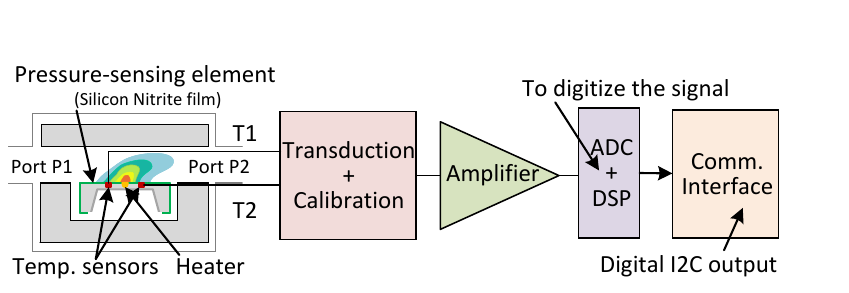}
\vspace{-1.0em}
\caption{Components of a DPS. }
\label{fig:thermalDPSComp}
\end{figure}

\subsection{\revision{DPS's Critical Role in Controlled Environments}}
DPS are widely used in applications that require monitoring the pressure difference between two points in a system. 
These sensors regulate airflow and maintain specific pressure levels within controlled environments. 
DPS offers high sensitivity, accuracy, and reliability.

\revision{While DPS find applications in various sectors, their role is particularly critical in the semiconductor industry, where they are essential for maintaining cleanroom integrity.}
Cleanrooms in semiconductor manufacturing serve as highly controlled environments designed to minimize airborne contaminants ~\cite{cooper1986particulate, kitajima1997requirements}.
\revision{Recent contamination incidents at major companies like Samsung and TSMC, resulting in losses exceeding \$1 billion~\cite{shilov_2019, yap_2018}, underscore the critical nature of maintaining cleanroom integrity.}

Besides cleanrooms, DPSs are commonly used in HVAC systems for buildings, offices, and hotels. Monitoring and controlling airflow enable efficient temperature regulation, air quality management, and energy optimization. DPSs are also employed in smart home systems, where they contribute to creating a comfortable and energy-efficient living environment.
DPSs are also used in healthcare settings, commonly in Negative Pressure Rooms (NPRs), to measure the negative pressure in the facility~\cite{miller2017implementing}. 

\revision{To contextualize our BaroVox attack, we examine the deployment of DPSs and sound systems in semiconductor cleanrooms as an example of a secure pressure-regulated facilities. 
The following sections dissect the cleanroom components pertinent to this attack, focusing on the integration of pressure sensors and audio systems. }

\subsection{{Components of a real-world cleanroom}}

\revision{Cleanrooms may vary in design and specifications across different organizations. 
However, the primary objective remains consistent: to prevent contamination by maintaining a sterile environment. 
Fig.~\ref{fig:cleanroom_diag} illustrates a standard cleanroom design.
The process begins with intake vents that channel outdoor air into the HVAC system. 
This system filters and conditions air by adjusting temperature and humidity, which are crucial for semiconductor production. Air pumps and compressors then push the air through HEPA filters, which remove contaminants, ensuring a clean environment. 
The RPM system is instrumental in constantly overseeing the pressure differences across cleanroom areas. 
The Room Pressure Monitoring (RPM) feeds data from the DPS into the HVAC, directing adjustments in airflow. 
The ensuing subsections spotlight the deployment of DPSs and sound systems within cleanrooms, given their significance to BaroVox. }

\begin{figure}[hb!]
    \centering
    \includegraphics[trim=1pt 1pt 1pt 1pt,clip,width=0.9\columnwidth]{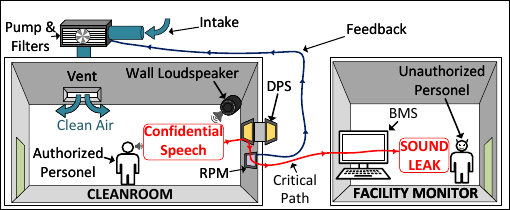}
    \caption{\revision{Components of a semiconductor cleanroom.}}
    \label{fig:cleanroom_diag}
    \vspace{-0.50em}
\end{figure}

\subsubsection{\revision{Sound Systems in Cleanrooms and Their Deployment}}
\revision{
Cleanroom operations \textit{demand} robust communication for coordination and safety adherence. 
Consequently, these environments often feature comprehensive sound systems~\cite{standard2004cleanrooms}. 
Intercoms serve a dual purpose of ensuring clear communication while reducing contamination risk. 
Additionally, cleanroom personnel often use protective clothing with integrated microphones, enabling efficient communication without removing protective gear, thus preserving the controlled environment. 
Wall-mounted speakers are strategically placed to broadcast announcements and facilitate communications. }

\revision{
These audio systems, crucial for operational efficiency, are frequently installed in close proximity to DPS due to space constraints and the need for integrated environmental control and communication. 
Moreover, the audio systems are channels for disseminating \textbf{\textit{proprietary and confidential data}} --- encompassing \textit{\textbf{process parameters, recipes, design information, and quality control protocols}}.
}

\subsubsection{\revision{DPS Deployment in Cleanrooms}}
\revision{DPSs are typically mounted in the cleanroom walls or close to the Building Management System (BMS). 
Each DPS employs a dual-port system connected to sampling tubes, strategically positioned to monitor pressure differentials between the cleanroom and adjacent spaces. 
One sensor port often connects to the cleanroom interior, while the other links to a reference point, usually via sampling tubes, enabling accurate pressure differential measurements. 
The RPM system utilizes these DPS to continuously monitor cleanroom pressure, transmitting readings to the BMS.
}

\subsection{Privacy implications of DPS deployment}
In this research, we show that the deployment of DPS in close proximity to audio systems, such as intercoms and speakers, raises significant privacy concerns. 
While this co-location is often necessary for facilitating communication~\cite{standard2004cleanrooms, litheaudio2023, zenitel2023}, it inadvertently creates an acoustic side-channel that can be exploited by attackers. 

The high sensitivity of DPS to pressure variations makes them vulnerable to unintentional acoustic coupling. Sound waves from nearby audio systems can induce minute vibrations on the sensor's diaphragm, causing measurable changes in the DPS output. Consequently, the DPS effectively functions as a makeshift microphone, allowing attackers to analyze the output variations and partially reconstruct the original audio signal. This enables eavesdropping on confidential conversations.

The privacy implications of this vulnerability are particularly concerning in environments where confidential discussions occur, such as cleanrooms, corporate meetings, healthcare consultations, or personal conversations in residential settings. Attackers may leverage this to gather valuable intelligence, compromise trade secrets, or invade personal privacy, all without the need for physical intrusion.
\revision{ Amid the rising IP war, this can be costly with IP and national security at stake.
Privacy implications of BaroVox in cleanrooms are discussed in detail in Appx.~\ref{appendix:cleanroombasics}.
}

\section{Related work}
In this section, we comparatively discuss BaroVox with state-of-the-art works in the following two categories.

{\textbf{Side-channel speech eavesdropping:}} Side-channel eavesdropping has been considerably studied in academic literature~\cite{michalevsky2014gyrophone, zhang2015accelword, anand2018speechless, anand2021spearphone, han2017pitchln, 7878599, marquardt2011sp,9517289, 9095984, cpams}. Several works, such as \cite{michalevsky2014gyrophone, zhang2015accelword, anand2018speechless, anand2021spearphone, han2017pitchln, marquardt2011sp, ba2020learning, wang2022mmphone, gao2022device, gao2022kite}, have investigated the vulnerabilities of motion sensor eavesdropping, most notably in mobile devices.
Sami et al. \cite{sami2020spying} devised a novel acoustic side-channel threat using the lidar sensors found in consumer-grade robot vacuums. 
Roy et al. \cite{roy2016listening} looked into the practicality of using a mobile device's vibration motor as a sound sensor. Nassi et al. \cite{nassi2022lamphone} exploit the vibrations of a hanging bulb inside a room to retrieve sound from desk lamp light bulbs through an optical side-channel attack. 
\cite{kwong2019hard,long2023side,nassi2021glowworm,nassi2023little,nassi2022little,wang2022wavesdropper,basak2022mmspy} are other works that focus on acoustic signal eavesdropping. 
These studies raised public awareness of the feasibility of recovering sound by analyzing non-acoustic data. Our work takes a unique approach to side-channel attacks by demonstrating the \textit{\textbf{first-ever use of this technique on pressure sensors}}, adding a new dimension to this field of study. 

\textbf{Attacks on pressure sensors:}
Tu et al. \cite{tu2021transduction} reveal threats of Electromagnetic interference spoofing attacks on tire pressure sensors to over/under-inflate car tire.
Barua et al. designed malicious music to create resonance in pressure sensors to fool the pressure sensor used in the RPM systems of a negative pressure room (NPR) to turn NPR's negative pressure into positive one~\cite{barua2022wolf}.   Our research distinguishes itself by highlighting the sensitivity of pressure sensors across a range of frequencies beyond the resonant frequency. Moreover, while both papers engage DPS sensors, they diverge significantly in focus and methodology. 
Barua et al. exploit resonant frequencies to manipulate pressure readings and create hazardous conditions --- an integrity attack. In contrast, our work reveals the potential of DPS as a covert channel for leaking acoustic information 
 --- a confidentiality attack. Additionally, we demonstrate the potential to extract sound signals from minor fluctuations in pressure readings, a new avenue for exploiting pressure sensors.



\section{Threat Model}

\begin{figure*}[!t]
\centering
\includegraphics[trim=2pt 10pt 2pt 2pt,clip,width=0.7\textwidth, height=0.1655\textheight]{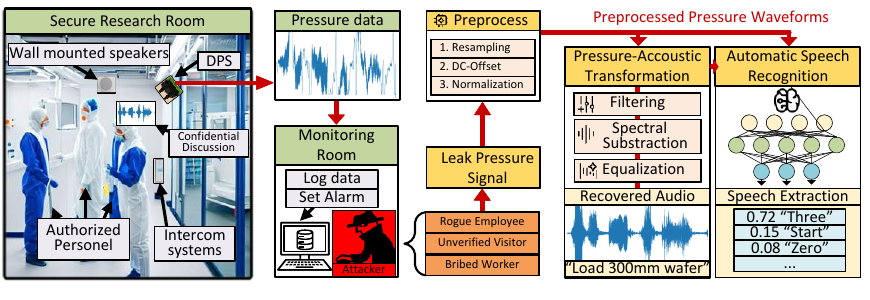}
\vspace{-0.50em}
\caption{Overview of the attack model.}
\label{fig:attack_model}
\vspace{-1.00em}
\end{figure*}

Fig.~\ref{fig:attack_model} depicts an overview of our attack model, outlining the attacker's target system, goals, capabilities, and assumptions.

\textbf{Attacker's Target:}
We consider a scenario in which the attacker targets a sensitive conversation or meeting taking place in an environment equipped with DPSs, such as a corporate boardroom, a secure research facility, a cleanroom in a semiconductor manufacturing plant, or any pressure-regulated room. The participants in the conversation are unaware of the potential for eavesdropping through the DPSs and assume that their discussion is confidential.

\textbf{Attacker's Goal:}
The attacker aims to reconstruct speech stealthily or recover sensitive information from the environment by exploiting the acoustic side-channel vulnerability in DPSs. 
The attacker could use this data for malicious activities, including espionage, blackmail, and theft.

\textbf{Attacker's Capabilities:}
The attacker is assumed to know the type and characteristics of the DPS deployed in the target environment. This information can be obtained through technical specifications or by studying similar systems as discussed in Sec.~\ref{sec:visibility}. The attacker also possesses the necessary technical skills and resources to process and analyze the captured pressure data using signal processing and machine learning techniques.

\textbf{Attack Scenarios:}
BaroVox can be used for either \textit{targeted eavesdropping}, where the attacker targets specific individuals, conversations, or events, or \textit{broad-spectrum eavesdropping}, where the attacker indiscriminately monitors the environment to recover any valuable or sensitive audio content.

\revision{
These attacks can be executed across various settings. In industrial and corporate environments, attackers can intercept confidential discussions about trade secrets or strategic plans. The vulnerability extends to private spaces, where DPSs in smart home or hotel HVAC systems could capture personal conversations. In healthcare facilities, government buildings, or military installations, BaroVox could be used to eavesdrop on confidential patient-doctor conversations or gather classified intelligence. The attack's effectiveness may vary based on factors like DPS model and environmental conditions, but the range of potential scenarios underscores the critical need to address this acoustic side-channel vulnerability.
}
\remove{\textit{Industrial espionage:} In industrial settings, such as manufacturing plants or research facilities, DPSs are used for process control and monitoring. An attacker can employ BaroVox to eavesdrop on confidential discussions related to trade secrets, proprietary technologies, or production processes.

\textit{Corporate espionage:} An attacker may target DPSs in corporate offices or meeting rooms to eavesdrop on confidential discussions, negotiations, or strategic planning sessions.

\textit{Privacy invasion in residential or hotel settings:} In smart home environments or hotel rooms equipped with DPSs for HVAC control or air quality monitoring, an attacker can exploit the BaroVox vulnerability to listen to private conversations or capture intimate audio content without the occupants' awareness.

\textit{Eavesdropping in healthcare facilities:} DPSs are commonly used in healthcare settings, such as hospital rooms or medical laboratories, to maintain proper room pressurization. An attacker can leverage BaroVox to eavesdrop on confidential patient-doctor conversations or sensitive medical discussions.

\textit{Intelligence gathering in government or military facilities:} In high-security environments, such as government buildings or military installations, DPSs may be deployed for monitoring and controlling critical infrastructure. An attacker can utilize BaroVox to intercept classified discussions or gather intelligence.

The effectiveness of the BaroVox attack may vary depending on factors such as the specific DPS model, the distance between the audio source and the DPS, the acoustic properties of the environment, and the presence of background noise. Nevertheless, the wide range of potential attack scenarios emphasizes the need to address the acoustic side-channel vulnerability of DPSs.
}

{\textbf{Attacker's Access Level:} 
\label{subsection:accessLevel}We assume that the attacker does not have physical access to the targeted environment or the ability to tamper with the DPS hardware. Instead, the attacker relies on remote access to the pressure sensor readings, either through a compromised system, a malicious insider, or by intercepting the sensor data transmitted over a network.

Attackers may obtain pressure reading data by exploiting personnel with clearance to access pressure sensor logs but not necessarily the targeted environment itself, such as a rogue employee, visitor, or maintenance worker. Our attack model focuses on situations where security measures protecting pressure log data are less stringent than those within the targeted environment.

Attackers may also tamper with the pressure monitoring system during delivery or installation, enabling it to transmit data via the Interne. 
\revision{Modern RPMs often have internet modules that allow remote connections, which can be exploited by the attacker to intercept data without being physically present~\cite{primex2021}}.
Attackers may also access archived pressure sensor logs, which are often overlooked but preserved for diagnostics or maintenance. Studies have shown that attackers exploit vulnerabilities in log security, using techniques like mapping sensor locations through security alerts or bypassing log protections~\cite{lee2019securely, paccagnella2020logging}.

\textbf{Assumptions:}
{We assume that the DPSs in the target environment are deployed in proximity to audio sources, such as speakers, intercoms, or areas where conversations occur. This assumption is based on the common practice of integrating audio systems with DPSs for various purposes}.

\section{Feasibility Study}
\label{sec:visibility}
Exploiting DPSs' acoustic side-channel vulnerability to recover speech signals requires a thorough understanding of the sensor's behavior, sampling rate, and frequency response. 
Sensor datasheets often conceal these details, which presents challenges for attackers attempting to reconstruct audio from pressure readings. Our analysis aims to highlight these challenges and demonstrate the attack's feasibility.

\begin{figure}[t!]
\centering
\includegraphics[trim={1px 1px 1px 1px},clip,width=0.75\linewidth]{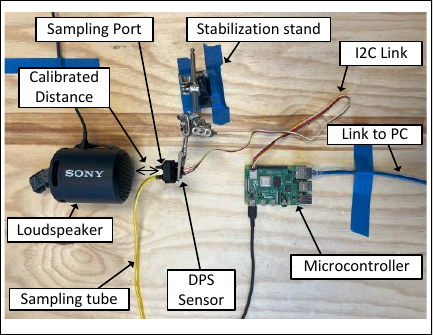}
\vspace{-0.5em}
  \caption{Our experimental setup for a feasibility study.}
  \label{fig:experiment}
  \vspace{-0.8em}
\end{figure}

\subsection{Experimental setup}
\label{subsection:experimental_setup}
The experimental setup is depicted in Fig. \ref{fig:experiment}. To showcase our attack's feasibility, we use the SDP800 DPS~\cite{datasheetsdp} (see~\ref{subsection:physicsofDPS}), securely mounted to avoid movement-related noise. 
A Sony XB13 \cite{xb13} speaker is placed 5 cm directly underneath one of the pressure ports to play the audio clips. 
To prevent sound signals from influencing the pressure in the other port, we connect a sampling tube to it, isolated by positioning its opening a meter away, shielded by a hard surface. 
Signals from the sensor are read using a Raspberry Pi 3 Model B~\cite{rs-online}.
We employ a custom Python script to convert the pressure readings into a format an attacker can process. 

\begin{figure}[b]
\centering
\includegraphics[trim={10px 22px 10px 18px},clip,width=0.49\textwidth]{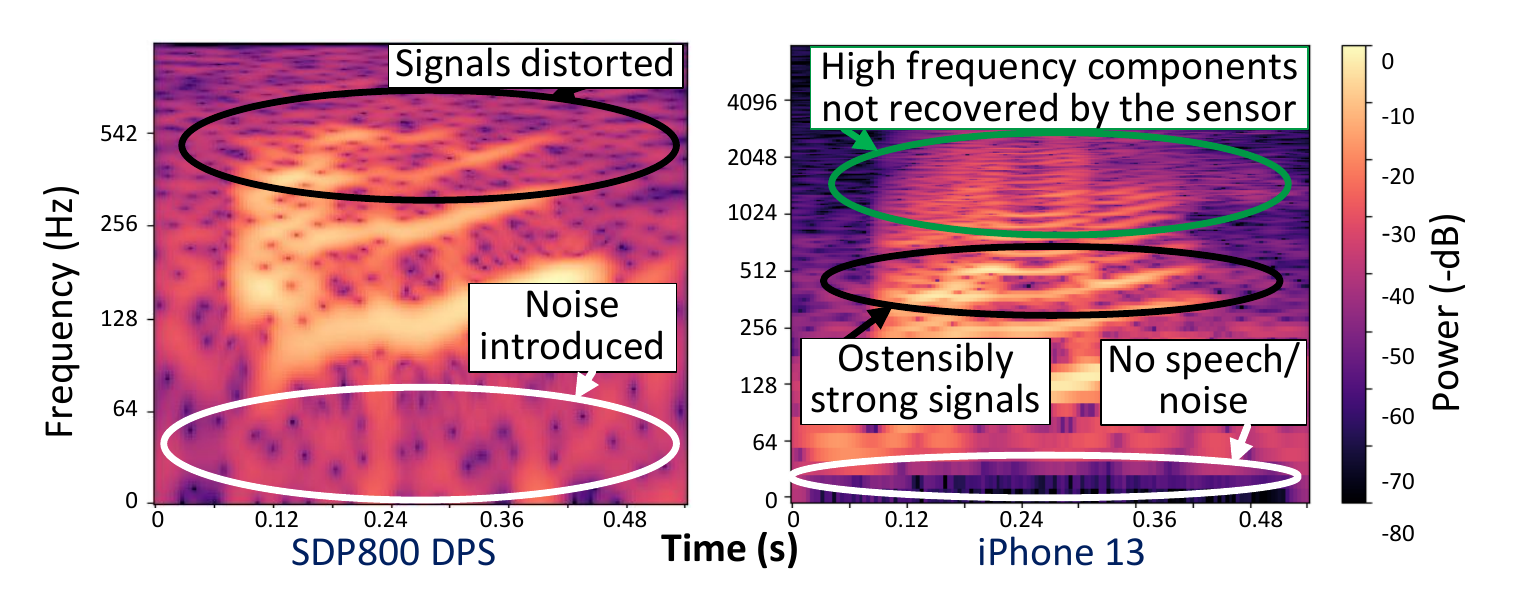}
  \vspace{-01.60em}
  \caption{STFT plot of a spoken word "one".}
  \label{fig:one_stft}
\vspace{-0.800em}
\end{figure}

\subsection{Primary observation}
Fig.~\ref{fig:one_stft} compares the Short-Time Fourier Transform (STFT) spectrum of the word "one" captured by an iPhone 13 microphone (right) and the SDP800 DPS (left). 
The spectrum demonstrates the variations in power across different audio frequency components over time.
Notably, the DPS's response is discernibly feebler and less extensive than the microphone’s, especially in frequencies surpassing $0.4$ kHz. 
Distortions are observable even in zones of ostensibly strong signals ($0.4$ - $0.85$ kHz). 
Furthermore, the DPS fails to register signals exceeding $0.9$ kHz. 
A comparative analysis of the Fast Fourier Transform (FFT) plots in Fig.~\ref{fig:one_FFT} echoes these observations, with the DPS showing subpar responsiveness to signals above $0.4$ kHz.
{This indicates a notable loss in acoustic fidelity when reconstructing speech from the DPS data (also see Fig.~\ref{fig:nonLinearity}).} 

\begin{figure}[h]
\centering
\includegraphics[trim={5px 20px 5px 20px},clip,width=0.41\textwidth]{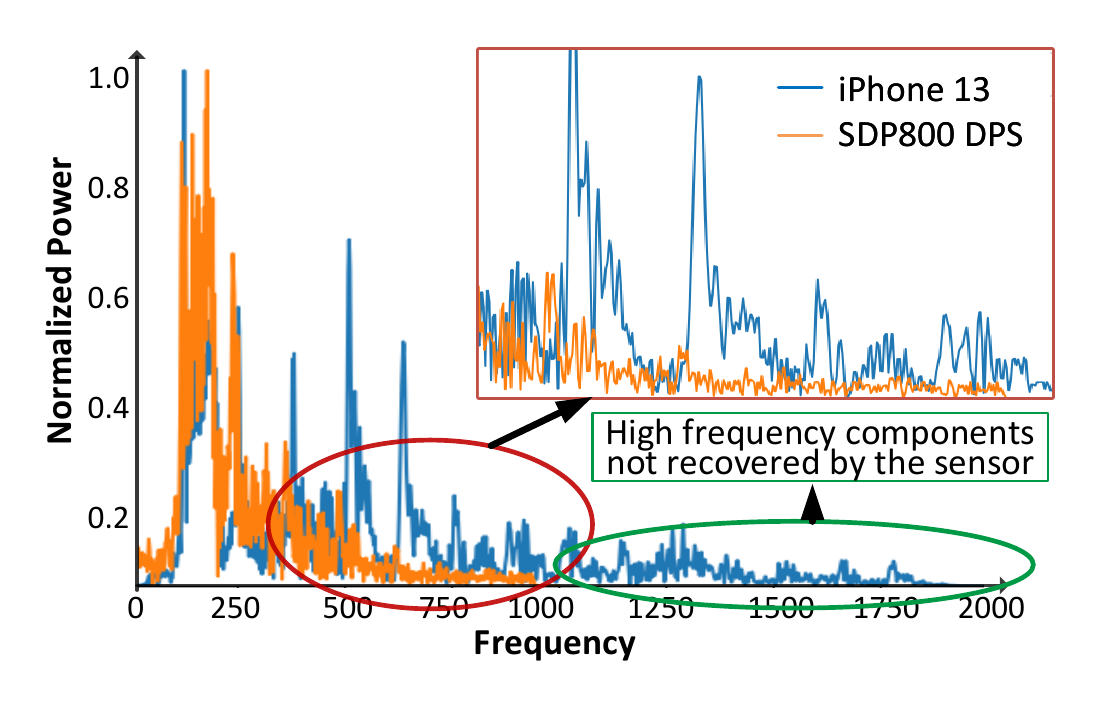}
\vspace{-0.5em}
  \caption{FFT plot of a spoken word "one".}
  \label{fig:one_FFT}
\vspace{-01.0em}
\end{figure}

\subsection{Challenges}
\label{subsection:challenges}
A successful attack requires recovery of information lost from pressure readings, at least partially.
Here, we examine the challenges attackers may face during this process.

\subsubsection{\textbf{Low sampling rate}} 
The sensor's low sampling rate is a significant challenge in executing the BaroVox attack.
For example, the SDP800 sensor used in our experiment has a theoretical maximum sampling rate of 2.2 kHz~\cite{datasheetsdp}.
However, as demonstrated in Fig.~\ref{fig:one_FFT}, it only captures frequencies up to approximately 0.9 kHz, with notably weak signals within the 0.6 - 0.9 kHz range. 
Consequently, given Nyquist’s theorem~\cite{vaidyanathan2001generalizations}, the sensor's effective sampling rate is approximated at 1.8 kHz, a shortfall of 0.4 kHz from its potential. 
This limitation restricts speech recovery within the sub-0.9 kHz range, significantly below the 4 kHz threshold necessary for intelligible speech~\cite{heide1998speech, villchur1973signal}, causing aliasing and complicating the recovery stage.

\subsubsection{\textbf{Non-linear frequency response}}
\label{subsection:non-linearity}

DPS can be depicted as a second-order dynamic system since it employs an elastic diaphragm for pressure force collection~\cite{whitmore2009improved}.
This leads to a non-linear frequency response to sound waves of varying frequencies.
To investigate this, we construct an audio file with a sine sweep wave ranging from 1 Hz to 2 kHz. We play the audio file through a speaker using the setup specified in Sec. \ref{subsection:experimental_setup}. We then reconstruct audio from pressure reading and analyze the data using FFT plots.
FFT plots of the reconstructed audio (Fig. ~\ref{fig:nonLinearity}) show reduced signal strength across frequencies compared to the original. 
{The original audio has consistent power up to 2 kHz, whereas the reconstructed signal loses power as frequency increases, dropping sharply after 0.4 kHz. }
At frequencies above 0.65 kHz, its power nears mere noise levels.
These insights can be leveraged to equalize the signal by boosting the strength of the affected frequencies (see Sec.\ref{subsubsection:equalization}).

\subsubsection{\textbf{Low Signal-to-Noise Ratio (SNR)}}
\label{subsection:lowSNR} 
DPSs yield non-zero readings even in the absence of sound due to the presence of ambient noise and pressure fluctuations.
Acoustic disturbances like machinery operation, air conditioning noise, and door activities further affect the readings.
These ambient noises introduce unwanted interference to the sensor's data, lowering the SNR and degrading the quality of reconstructed speech.
To mitigate this issue, we applied filters and spectral subtraction techniques to improve the audio recovery quality (see Sec.~\ref{subsection:DS-1}).

\section{{Attack Design and Implementation}}

This section outlines the attack system design.
We begin with a mathematical model to demonstrate sound waves' impact on DPSs and introduce two BaroVox design solutions. 

\begin{figure}[t]
\centering
\includegraphics[trim={5px 20px 5px 35px},clip,width=0.445\textwidth]{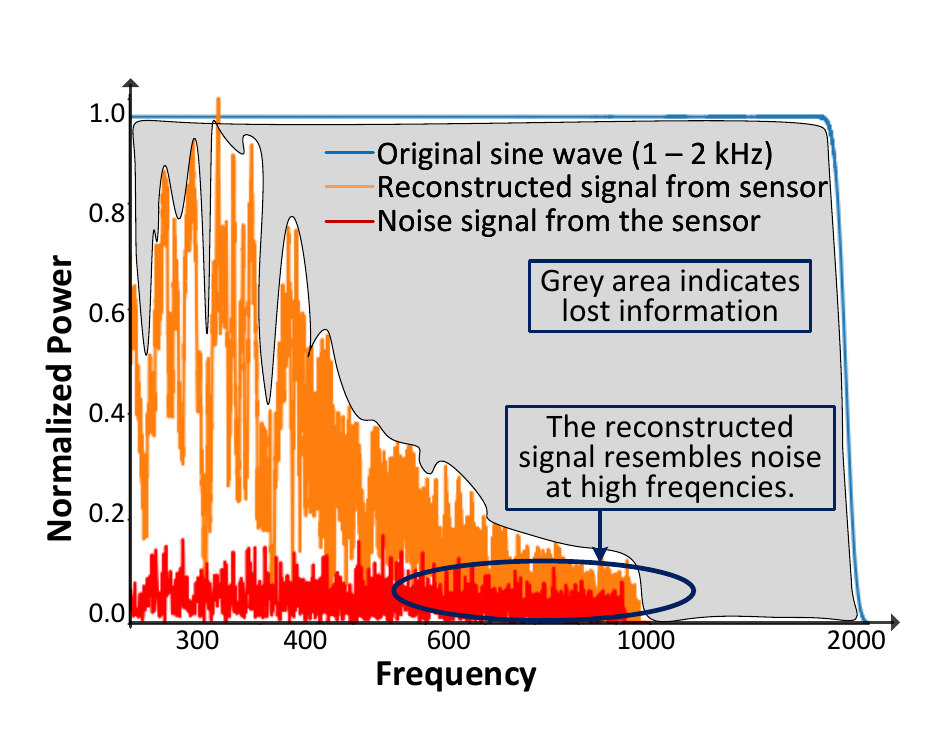}
\caption{Non-linear frequency response of the SDP800 DPS for a sine sweep wave with a frequency range of 1 to 2 kHz. }
\label{fig:nonLinearity}
\end{figure}
\subsection{Modeling effects of sound on DPSs}
\label{subsec: Modeling Sound effects on DPSs}

Sound waves, as disturbances propagating through matter, create varying local pressures by compressing and expanding air. Thus, sound can be modeled as a pressure wave. 
If a sound is played at a frequency $f$ with an initial phase $\phi$ and a wavelength of $\lambda$, then the change in pressure due to sound ($\Delta P_s(t)$) at a distance $x$ from the sound source can be represented: 
\begin{equation}
\begin{aligned}
\Delta P_s(t, x) &= A(x, f, \mu) \cdot P_{smax}\cdot sin(kx \pm f t + \phi )
\label{eqn:sound_as_pressure}
\end{aligned}
\end{equation}
\noindent
where $P_{smax}$ is the maximum pressure change due to sound and $A(x, f)$ represents the attenuation of a sound wave, which depends on distance $x$ and frequency $f$ of the audio source and noise $\mu$.

If there is a DPS at a distance $x=x_0$ from a sound source, we model the perturbation in the reading of the DPS as a result of a sound wave as follows:
\begin{equation}
P(t) = P_o(t) + \Delta P_s(t, x_0)
\label{eqn:sound_effect}
\end{equation}
\noindent
where $P(t)$ is the total measured pressure by the DPS, and $P_o(t)$ is the original pressure and noise read by the DPS without sound. 
\toremove{A visual representation of sound's effect on DPSs is provided in Appx.~\ref{appendix:modelingsound}.
The subsequent sections discuss designs allowing attackers to translate pressure readings into sound waves.}

\begin{figure*}[ht]
\centering
\includegraphics[trim={5px 33px 5px 30px},clip,width=1.01\textwidth]{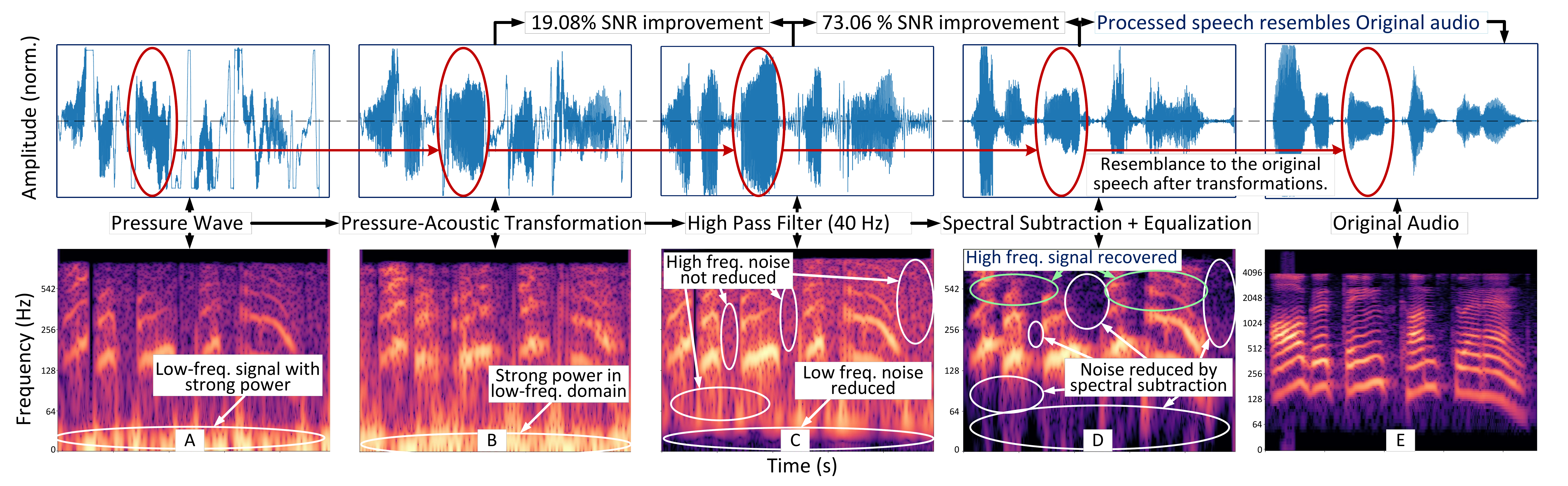}
\vspace{-0.99000em}
\caption{System Design I: Design overview and its effect on a speech recorded by a pressure sensor.}

\label{fig:systemDesignOne}
\vspace{-0.90000em}
\end{figure*}
\subsection{Design solutions: overview}
\label{subsec:ds-overview}
BaroVox offers two design approaches to address challenges discussed in Sec.~\ref{subsection:challenges}.
Both designs pivot on a Pressure-Acoustic Transformation (PAT) outlined below. 
Fig.~\ref{fig:attack_model} summarizes how these designs integrate into the attack model.

\textbf{Pressure-Acoustic Transformation (PAT)}:
PAT converts DPS readings into wave files.
PAT reduces the influence of the normal differential pressure of the target environment using Eqn.~\ref{eqn:sound_recov}. 

\vspace{-0.90000em}
\begin{equation}
S(t) = P(t) - \bar{P}(t)
\label{eqn:sound_recov}
\end{equation}

\noindent
where $S(t)$ is the amplitude of the sound wave at time $t$, $P(t)$ is the instantaneous pressure reading from the DPS and $\bar{P}(t)$ is the mean value of $P(t)$. 
This levels the DC offset due to the normal differential pressure of the room by subtracting the mean of the pressure signal $(\bar{P}(t))$ from $P(t)$.
$S(t)$ is further refined using either of the design solutions discussed below.

\textbf{Design Solution I (DS-I):}
\textbf{DS-I} aims to enhance the SNR of reconstructed speech ($S(t)$) using a series of cascaded digital signal processing techniques for improved audio quality.
It employs a smart spectral subtraction method in conjunction with standard normalization, high-pass filtering, and DC-offset reduction techniques. 
This approach proves valuable for attackers when the retrieved speech is substantially distorted by noise, efficiently mitigating noise and other audio signal distortions.
Furthermore, DS-I is not bound by vocabulary size, providing flexibility in recognizing a wide range of words and phrases.
Sec.~\ref{subsection:DS-1} discusses \textbf{DS-I} in detail.

\textbf{Design Solution II (DS-II):}  \textbf{DS-II} develops an Automated Speech Recognition system employing Deep Learning. It utilizes SpeechCommands datasets and utilizes a classification model.
{Attackers may prefer DS-II when more precision is desired and focus is required on specific critical words, numbers, or instructions in the target environment setting.}
\toremove{In the subsequent sections, we will delve into both design solutions, starting with DS-I. }

\subsection{Design solution I (DS-I)}
\label{subsection:DS-1}
Fig.~\ref{fig:systemDesignOne} provides a comprehensive overview of the techniques used in DS-I. 
Below, we delve into a deeper analysis of each technique.

\subsubsection{\textbf{High-pass filter}} 
A high-pass filter refines the recovered audio ($S(t)$) by enhancing its SNR. 
We initially acquire samples from the DPS without sound to observe acoustical disturbances induced by various environmental conditions in the target environment.
The FFT of this noise data, shown in Fig.~\ref{fig:nonLinearity} (red line), displays dispersed noise energy across frequency components. 
However, the speech recorded by an iPhone 13 in Fig.~\ref{fig:one_stft} (right) does not contain speech data in frequencies less than $40$ Hz. 
In contrast, the audio from the SDP800 DPS in Fig.~\ref{fig:one_stft} (left) is filled with noise below $40$ Hz, which is undesirable. 
Therefore, we apply a 3rd-order Butterworth high-pass filter with a cutoff frequency of $40$ Hz to remove the low-frequency component of the noise.
If not removed fully during PAT (see Eqn.~\ref{eqn:sound_recov}), the DC offset generated by the ambient pressure in the room would likewise be nullified using the filter. 

To further study the filter's influence,  we record a short three-second speech and reconstructed the audio from the sensor data using Eqn.~\ref{eqn:sound_recov}.
The speech reads: "\textit{All good things come to an end.}" 
The resulting SNR value of the reconstructed speech is $7.464~dB$. With the high pass filter, SNR rises to $8.888~dB$ -- 19.08\% improvement.
This effect is evident in Fig.~\ref{fig:systemDesignOne} (C)'s spectrum and waveform plot. 
Post PAT transformation from pressure wave (A) to speech data (B), prominent power in the low-frequency domain emerges. However, after the high-pass filter application, this noise diminishes (C), making the waveform more congruent with the original audio.

Given the noise in other frequency ranges overlaps with speech data (see Fig.~\ref{fig:one_stft} (right)), a more sophisticated method to eliminate this is discussed in the upcoming subsection.

\subsubsection{\textbf{Spectral subtraction}}
\label{subsubsection:Spectral} 
Spectral subtraction subtracts an estimate of the noise spectrum from the speech spectrum to get a denoised spectrum. 
Direct subtraction of noise from speech is theoretically ideal but can distort elements of the speech signal with noise-like characteristics. 
To address this, we differentiate between two sound types within the speech signal: percussive and harmonic components, and their susceptibility to noise.

\textbf{Percussive components:} 
Characterized by their brief, non-steady nature, percussive sounds lack a clear pitch or tonal quality~\cite{ladefoged2014course}. Examples include consonants like plosives and non-pulmonic sounds produced by rapid, irregular vibrations within the vocal tract~\cite{gilman2019science}. With sharp attack and decay times, these components exhibit noise-like spectral characteristics.

\textbf{Harmonic components:} 
In contrast, harmonic components possess a clear, identifiable pitch or tone, produced by steady-state vibrations of the vocal cords~\cite{ladefoged2014course}. These components encompass vowels and voiced consonants like \textit{`l'}, \textit{`n'}, and \textit{`r'}, exhibiting periodic characteristics~\cite{moulines1995non}.

Given the noise-like attributes of percussive sounds, they are inherently vulnerable to distortion during spectral subtraction, impacting the speech signal's integrity~\cite{cano2014phase, tachibana2014harmonic}. 
Consequently, while employing spectral subtraction,\textit{ emphasis should be on mitigating noise within harmonic components}.
Subtraction should be \textit{cautiously applied to percussive elements to preserve crucial signal information}. 
Fig.~\ref{fig:harm_perc_filtering} in Appx~\ref{appendix:phs} illustrates the spectral subtraction process. Hereafter, we comprehensively analyze each part of the process separately.
\begin{algorithm}[ht!]
\setstretch{1}
	\footnotesize
	\DontPrintSemicolon
	\caption{PHS using median filtering.}
	\label{alg:MFAlgorithm}

\KwIn{$s$: Input audio, $W$: Window size, $H$: Hop size, $n\_iter$: Number of iterations}
\KwOut{$P$: Complex spectrogram of percussive component, \newline
$H$:  Complex spectrogram of the harmonic component}

$S[n\_iter,K] \gets$ STFT of $s$ with window size $W$ and hop size $H$, where N is the time frame index, and K is the frequency index\\
$S\_mag \gets$ magnitude spectrogram of $S$\\
$S\_phase \gets$ phase spectrogram of $S$

\For{$n = 1$ to $n\_iter$}{
Compute the harmonic component energy envelope:
$E_h[n, k] \gets \sum\limits_{j=1}^{\infty}S\_mag[n, j\cdot k]$ \\
Compute the percussive component energy envelope:
$E_p[n, k] \gets \max\limits_{j\neq 0, j\cdot k\leq K} S\_mag[n, j\cdot k]$; \\
Compute the percussive mask: 
$M_p[n, k] \gets \frac{E_p[n, k]}{E_h[n, k] + E_p[n, k]}$ \\
Compute the harmonic mask:
$M_h[n, k] \gets \frac{E_h[n, k]}{E_h[n, k] + E_p[n, k]}$\\
Compute the complex spectrogram of the percussive component: 
$P[n, k] \gets M_p[n, k] \cdot S[n, k] \cdot~exp(i \cdot S\_phase[n, k])$ \\
Compute the complex spectrogram of the harmonic component: 
$H[n, k] \gets M_h[n, k] \cdot S[n, k] \cdot~exp(i \cdot S\_phase[n, k])$\\
}

\Return{$P$, $H$}
\end{algorithm}

\textbf{Percussive-Harmonic Separation (PHS)}: PHS is the initial step in spectral subtraction (see Fig. \ref{fig:harm_perc_filtering} (B)). We use a technique based on median filtering to separate the speech signal into harmonic and percussive components~\cite{fitzgerald2010harmonic, driedger2014extending}. 
\textbf{Median filtering} is a signal processing technique that aims to remove noise from a signal by replacing each sample with the median value of a group of neighboring samples. 
Algorithm~\ref{alg:MFAlgorithm} shows the technique we used for PHS. 
First, the STFT of the signal is calculated \textit{(line 1)}, followed by computing the magnitude and phase of the spectra \textit{(lines 2-3)}.
Median filtering is then applied to the magnitude spectra across sequential frames, enhancing percussive components and suppressing harmonic ones~\textit{(line 5)}.
Subsequently, median filtering across frequency bins is performed on magnitude spectra to bolster harmonic components while suppressing percussive events~\textit{(line 6)}. 
We use the two resulting median-filtered spectrograms to generate masks~\textit{(lines 7-8)}.
These masks are applied to the original spectrogram to separate the harmonic and percussive parts of the signal~\textit{(lines 9-10)}. 
The algorithm yields two signals: one containing only percussive components and one containing only harmonic components~\textit{(line 11)} (also see Fig.~\ref{fig:harm_perc_filtering} (B)).

\textbf{Characterizing noise:} 
Before implementing noise reduction, it is vital to model the noise properties~\cite{zhang2021influence, nishimura2004noise}. 
To characterize the statistical features of noise, we capture a segment of noise in the target environment and produce its power spectrum.

\textbf{Spectral subtraction}: We then apply spectral subtraction only to the harmonic signal using an estimate of the residual noise spectrum obtained above (see Fig. \ref{fig:harm_perc_filtering} (C)). 
This step results in an enhanced harmonic signal with minimized noise artifacts. 
Prior to applying spectral subtraction to the percussive component, the residual noise spectrum is downscaled to prevent compromising the speech signal quality. 
The optimal downscale factor is determined through subjective listening tests. 
Subsequently, a denoised speech signal is reconstructed by combining the adjusted harmonic and percussive signals. 
\toremove{Fig.~\ref{fig:harm_perc_filtering}  in Appx~\ref{appendix:phs} visualizes an example of the spectral subtraction process facilitated by the PHS process.}
Fig.~\ref{fig:systemDesignOne} (D) demonstrates the amalgamated effect of spectral subtraction and equalization (see Sec.~\ref{subsubsection:equalization}) on the spectrum and waveform plot, highlighting the noise reduction across various frequency ranges.
Notably, applying spectral subtraction alone leads to an $11.6\%$ increase in the SNR of the reconstructed speech, improving it from $8.888$ to $9.921~dB$. The ensuing section will address the final step of DS-I: equalization.

\subsubsection{\textbf{Equalization}}
\label{subsubsection:equalization} 
As noted in Sec.~\ref{subsection:non-linearity}, DPS exhibits a non-linear frequency response. 
To rectify this, we implement equalization, aiming to establish a desired tonal balance and sound quality. 
This process involves adjusting the gain of various frequency components of the signal to offset any deviations from a flat frequency response.
We use \textbf{frequency-domain equalization} technique to modify the sensor's frequency response. 
This technique efficiently amplifies or diminishes specific frequency ranges without influencing other portions of the signal~\cite{li2020time}.

\begin{figure}[t]
\centering
\includegraphics[trim={8px 10px 8px 25px},clip,width=0.46\textwidth]{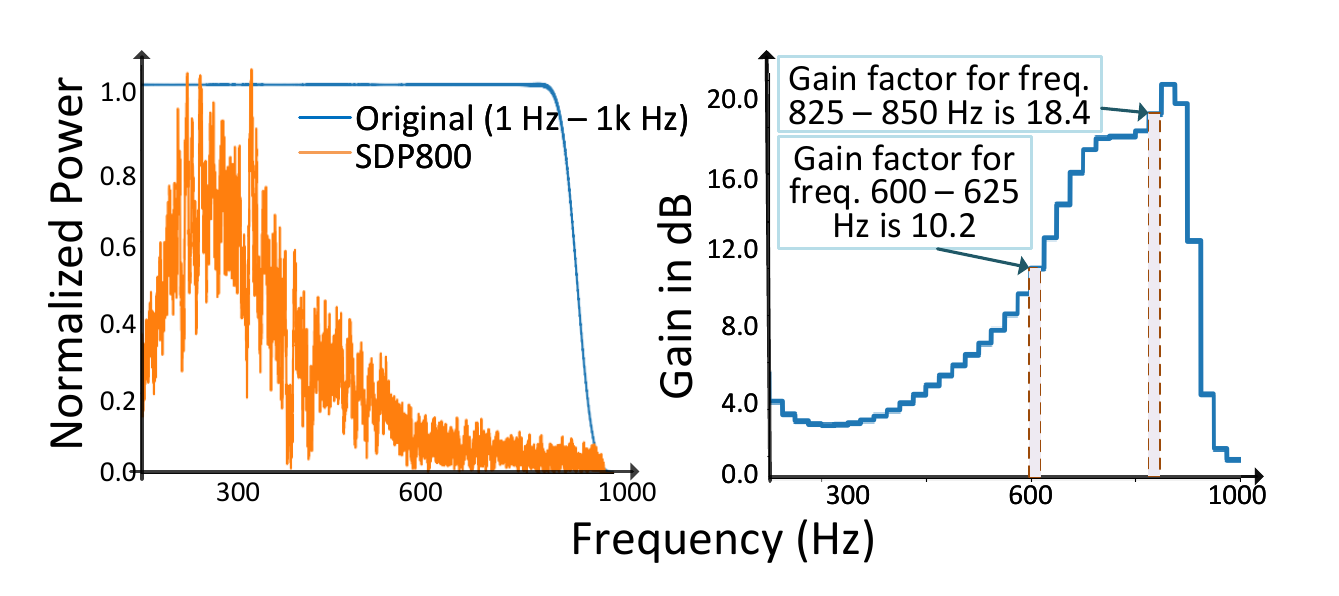}
\vspace{-0.80em}
\caption{(Left) Frequency response of the SDP800 DPS for a sine sweep wave (1 Hz - 2 kHz). (Right) The gain factor for each frequency band.}
\label{fig:equalization}
 \vspace{-1.01000em}
\end{figure}

\textbf{Frequency-Domain Equalizer (FDE):} We design our FDE using a bank of bandpass filters. 
The design process involves choosing the right number of filters, setting their center frequencies and bandwidths, and defining the necessary gain or attenuation for each frequency~\cite{falconer2002frequency}.
An experiment was conducted to define these parameters: a sine wave with frequencies ranging from 1 to 1 kHz was played and recorded using the SDP800 DPS. 
Normalized FFT plots of both the original and the DPS-reconstructed signals are presented in Fig.~\ref{fig:equalization} (left). 
It is clear that the sensor's non-linear response affected the reconstructed signal. 

\textbf{Filter specifications}: 
The number of bandpass filters required is determined by various parameters, including the signal’s frequency range (1 kHz in our case), desired frequency resolution, and the filter bank's complexity.
A balance between filter bank complexity and required frequency resolution is needed for optimal computational efficiency and equalization accuracy. 
Our approach uses 40 bandpass filters, each with 25 Hz equal-width bands.
Subjective tests indicated that increasing the filter count did not significantly improve sound quality, guiding our decision for $40$ filters.

To calculate the gain at each frequency band, we used the reconstructed signal and the original sine wave. We put them through Fourier transformation to get their frequency spectrums. The gain for each band is computed as the ratio of the original sine wave's amplitude to the amplitude of the reconstructed signal at the band's center frequency. Fig.~\ref{fig:equalization} depicts the gain factor for each frequency band. These gain factors were then applied to the output signal for each frequency band to produce an equalized signal that balanced the non-linear response of the DPS. 
Fig.~\ref{fig:systemDesignOne} depicts the result of the equalization process, in which the higher frequencies of the input signal are amplified after passing through the equalizer, increasing the SNR value by $55.04\%$ from $9.921~dB$ to $15.382~dB$.

\subsection{Design solution II (DS-II)}
\label{subsection:DS-2}
DS-I offers enhanced audio with a lower SNR, enabling attackers to perform partial manual speech recognition.
However, there are times when the attacker wants to get a more precise and accurate recognition of words spoken inside the target environment.
For this purpose, we develop an Automated Speech Recognition (ASR) DL model. The ASR solution classifies the pressure signals into their textual representations. 
\toremove{Subsequent sections detail task and dataset selection, data collection methodology, and DL model architecture.}

\subsubsection{\textbf{Task and dataset selection}}
With DS-II in operation, the attacker aims to \textbf{automatically extract and categorize} spoken words from the captured pressure signals.
Given communication inside sensitive environments is often restricted for clarity and brevity, we want our model to learn to classify pressure wave signals into a limited set of \textbf{keywords}. 
For this, we focus on types of speech containing critical information; specifically, we target spoken digits and commands. 
For example, the sentence "Load 3.1-millimeter wafers" presents insight into both the manufacturing process and product specification within a semiconductor manufacturing cleanroom through the command "load" and the number "3.1," respectively. 
We focus on digit and command classification using the 
\textbf{SpeechCommands}~\cite{speechcommandsv2} dataset. SpeechCommands contains common speech commands from multiple speakers in various environments.
An overview of the datasets is available in Appx.~\ref{appendix:data}. 

\textbf{Considerations on vocabulary limitations:} While it is evident that real-world applications would present a more diverse vocabulary, the current task is focused on unveiling the latent risks tied to this unexplored side-channel attack. Employing a constrained vocabulary for this proof-of-concept might not entirely reflect the breadth of real-world scenarios, yet it aptly illustrates the potential of the proposed approach. Generating a comprehensive dataset is undoubtedly resource-intensive and costly. However, it is crucial to emphasize that the primary aim here is to spotlight potential security vulnerabilities. {In a scenario where confidential information is the target, it is reasonable to assume that attackers, understanding the value of the information at stake, would not balk at investing in a robust dataset to realize their illicit objectives.}

\subsubsection{\textbf{Dataset transformation}}
As the dataset involves perfect \textit{microphone} recordings of speech utterances, we use an \textbf{{acoustic-pressure transformation strategy}} to transform them into \textit{pressure-wave} datasets.
Given sound clips and keyword labels, denoted as $(x_i(t), y_i)$, we play the speech through a speaker in the proximity of a pressure sensor.  The data is collected in a research facility, with varying distances and orientations between the speaker and the pressure sensor, as discussed in Sec.~\ref{sec:evaluation}.
This creates a new dataset that pairs each sound signal $x_i(t)$ with the recorded pressure $P_i(t)$. 
We then process $P_i(t)$ by using \textbf{{PAT}} (see Sec.~\ref{subsec:ds-overview}) yielding the signal ${S_i}(t)$. We create a final dataset by pairing the reconstructed speech signal ${S_i}(t)$ with their original keyword labels, forming pairs $({S_i}(t), y_i)$.
An attacker can use this new dataset to train models.

\subsubsection{\textbf{ASR model architecture}} 
While numerous time series classification models are available in the literature~\cite{ts1, ts2, ts3}, not all of them are well-suited for our specific application due to the challenges posed by the low SNR and the low sampling rate of the sensor. 
We build upon \textit{ResNet}~\cite{resnet, audioresnet}, known for extracting information from sparse signals~\cite{receptivefields}.
Our contribution centers around our ASR model, designed to address the challenges posed by low SNR and non-linear frequency response of the DPS. This  includes \textit{learnable} \textbf{\textit{denoising autoencoder}} and \textbf{\textit{equalization layers}}, providing a robust solution to these specific issues.
It also employs a spectrogram representation defined by FFT bins, window length, and hop size parameters. 
We sweep these parameters to find the optimal receptive field for our transformed dataset. 
The full explanation of each spectrogram parameter and the model architecture is included in Appx.~\ref{appendix:models}.

\section{Evaluation}
\label{sec:evaluation}

This section evaluates BaroVox's design solutions across various metrics and scenarios. The experiments are conducted in a seminar room of an anonymous research lab, simulating real-world conditions while ensuring a controlled environment for data collection.
\toremove{First, we describe the methodologies and metrics employed to assess each design solution. Then, we present the experimental outcomes.}

\subsection{Methodology and metrics}

\subsubsection{\textbf{Manual Speech Recognition (MSR)}}
{
We evaluate DS-I using MSR. For this task, we recruited 18 volunteers from our institution. The survey includes people with diverse linguistic backgrounds, with only eight considering English as their primary language. Others speak languages from Africa, South Asia, and East Asia.
We use the setup discussed in Sec.~\ref{subsection:experimental_setup} to prepare the evaluation dataset.
We record 20 sentences focused on general conversations, scientific theories, and semiconductor fabrications using an iPhone 13 microphone and DPS. 
We use DS-1 to perform pressure-acoustic transformation, remove background noise, and improve the overall quality of the speech. 
We then use the \textbf{word error rate (WER)} and the \textbf{mean opinion score (MOS)} metrics for evaluation. 
For both metrics, each volunteer is kept in a quiet room for listening.

\textbf{WER}: WER is a standard measure for speech recognition tasks~\cite{errattahi2018automatic}. To calculate WER, we play the reconstructed audio data from the DPS separately for each volunteer and request them to transcribe it based on their comprehension. 
WER is computed by comparing the transcription to the actual spoken words and counting errors using the formula: $WER = (S + D + I) / N $.
$S$, $D$, and $I$ denote the number of incorrectly transcribed, missing, and incorrectly added words by the volunteer, while N represents the total number of words in the actual spoken recording. 
A lower WER indicates greater intelligibility of the reconstructed audio.

\textbf{MOS}: MOS serves as a subjective measure of the perceived speech \textit{quality}~\cite{leng2021mbnet}. 
MOS assesses how well volunteers can comprehend the content of the reconstructed speech. 
Participants are instructed to listen to the reconstructed speech first and then the original audio. 
Subsequently, they rate the content-wise similarity between the two on a scale of 1 to 5. 
For instance, if volunteers perceive that the reconstructed audio is understandable and resembles the original audio content, they assign a score of 5. 
Conversely, if they believe that the reconstructed speech differs significantly from the original speech, they assign a score of 1. 
This approach allows us to quantify how effectively our system reconstructs speech that is understandable and akin to the original content.
}

\subsubsection{\textbf{Automatic Speech Recognition (ASR)}}
{
We evaluate DS-II on automated classification tasks through the transformed SpeechCommands~\cite{speechcommandsv2} datasets. 
We mainly use \textbf{accuracy} as the evaluation metric to measure our model's ability to accurately and truthfully recognize the given classes. \toremove{A higher classification accuracy makes it easier for attackers to extract information reliably.}

\subsection{Results for MSR}
\label{subsec:MSRevaluation}

\subsubsection{\textbf{Performance on general speech}}
Table~\ref{table:generalMSR} in Appx.~\ref{appendix:performance} shows the average \textbf{WER} and \textbf{MOS} of each ground truth sentence over the responses of the 18 volunteers. 
On average, participants had a WER of \textbf{0.35}, and a MOS of \textbf{4.09/5}. A WER of 0.35 means volunteers can reconstruct more than 60\% of the speech effectively, a significant achievement from the attacker's perspective. 
A MOS of 4.09/5 also shows volunteers perceived a notable content resemblance between the reconstructed speech and the original audio. 
These results show that humans can identify and reconstruct everyday speech partially from pressure-wave signals.

{\subsubsection{\textbf{Performance on sensitive information}}
To assess the impact of targeted eavesdropping on sensitive information, we focus on sentences containing confidential data related to semiconductor manufacturing processes. A detailed discussion of the key components of a cleanroom, including the deployment of DPSs and sound systems, is provided in Appx.~\ref{appendix:cleanroombasics} to contextualize the BaroVox attack in such environments.
Table~\ref{table:semicondMSR} shows the aggregated responses of the 10 volunteers on the chosen sentences and the type of confidentiality breach each sentence target. The results are an average \textbf{WER} and \textbf{MOS} of \textbf{0.45} and \textbf{3.36/5}, respectively. These results confirm that humans can partially understand speech related to semiconductor fabrication from pressure-wave signals even without extensive field knowledge. The results are comparatively lower than the performance on general speech,  and this is due to the participants' limited familiarity with semiconductor cleanroom contexts. To validate that, we conduct the survey on the remaining 8 volunteers, but at this time, we provide them with a short text about the semiconductor manufacturing process. The content of the text is available in Appx.~\ref{appendix:semiconductorProcessDesc}. The survey results in an average \textbf{$\mathcal{WER}$} score of \textbf{0.29} and \textbf{$\mathcal{MOS}$} score of \textbf{3.71}. The score for each individual sentence is provided in Tab.~\ref{table:semicondMSR}. Given an attacker's presumed familiarity with the content in Appx.~\ref{appendix:semiconductorProcessDesc}, we infer that employing the BaroVox attack could enable them to accurately reconstruct over 70\% of cleanroom speech. \revision{In the semiconductor industry, even fragments of information about processes or specifications can be valuable to competitors, making this level of reconstruction particularly concerning.}

We demonstrate the effect of DS-I on some of the sentences that we used in the survey in the following link: 
\textcolor{blue}{\href{https://sites.google.com/view/barovox-a-fly-on-the-wall/home}{{BaroVox}}}.

\setlength{\tabcolsep}{2pt}
\begin{table}[t!]
		\footnotesize
		\centering

\begin{tabular}
{|p{4.5cm}|L{1.1cm}|L{1.1cm}|p{1.1cm}|}
\hline
\graycell \textbf{Ground Truth Sentence} & \graycell \textbf{WER /$\mathcal{WER}^1$}&\graycell{ \textbf{MOS /$\mathcal{MOS}^1$}} & \graycell{ \textbf{Breach Type$^2$}} \\ 
\hline
Run the diffusion process at three hundred kelvin for nine minutes &  0.38/0.20 & 3.45/3.50 & TC\\ \hline
Use process B for the trench isolation & 0.62/0.32 & 3.00/3.88 & TC \\ \hline
 Etch the pattern using SF6 plasma &   0.40/0.25 & 3.55/3.25 & TC\\ \hline
Sputter a one hundred and fifty nano meter layer of Aluminum copper alloy for contacts & 0.48/0.45 & 3.00/3.38 & TC \\ \hline
The defect rate for lot number seven is two percent & 0.30/0.15 & 3.27/3.88 & TC\&QC\\ \hline
Overheating issues impacted four percent of the latest batch & 0.28/0.23 & 3.64/4.00 & QC \\ \hline
Decrease metal layer flowrate & 0.67/0.65 & 3.00/2.75 & TC\\ \hline
Use different etching gas for oxide layer & 0.40/0.29 & 3.36/3.50 & TC \\ \hline
Perform a defect analysis on the wafer batch &  0.39/0.29 & 3.55/3.50 & TC\&QC \\ \hline
Reduce the concentration of the cleaning solution. & 0.53/0.29 & 3.82/4.50 & TC\&QC \\ \hline
 \textbf{Average}& 0.45/0.29 & 3.36/3.71 & \\ \hline
\multicolumn{4}{l}{${\textbf{1:}}$ $\mathcal{WER}$ \& $\mathcal{MOS}$: Volunteers' scores after receiving context.} \\\multicolumn{4}{l}{${\textbf{2:}}$ TC - Trade secret, QC - Quality control} \\
\multicolumn{4}{l}{${\textbf{NB:}}$  IRB exemption approval obtained to conduct the survey.} \\
\end{tabular}
{\centering\caption{\label{table:semicondMSR}Performance on semiconductor-focused sentences.}}
 \vspace{-01.60em}

\end{table}

\subsection{Results for ASR}

\begin{table}[b]
    
		\footnotesize
  \centering
\vspace{-1.0em}
    \begin{tabular}{|c|c|c|c|c|}
         \hline
         \graycell \textbf{Models} &\graycell \textbf{FFT size} &\graycell  \textbf{Our ASR Model} &\graycell  \textbf{ResNet}  \\
         
         \hline
        SpeechCommand64 &  64 & 80.37\% &68.16 \\ \hline
        SpeechCommand128 &  128 & 86.82\% &70.35\\ \hline
        \textbf{SpeechCommand256} &  \textbf{256} & \textbf{90.51\%} &{80.14} \\ \hline 
    \end{tabular}
    \caption{\label{tab:ASRperf}Performance metrics of ASR.}
    \vspace{-1.0em}
\end{table}

\subsubsection{ \textbf{Performance on the testing dataset}}
Table~\ref{tab:ASRperf} demonstrates the performance of the model when trained across various FFT bin sizes, specifically 256, 128, and 64 bins. A clear correlation between the number of bins and performance is observed, with larger bins yielding improved results. In particular, models using 256 FFT bins (\textbf{SpeechCommand256}) outperform the others, achieving accuracy of {90.51}\%. This is not surprising since larger bins imply a more receptive field, which provides better frequency resolution. 
We compare our model's performance to others who conducted word classification on the original SpeechCommand dataset. We find that the current state-of-the-art ML technique achieves 98.32\% accuracy~\cite{gu2021efficiently}.
This difference is deemed acceptable, given the speech signals used in the original model have a higher sampling rate (16 kHz). 
To demonstrate the impact of our contribution --- the addition of \textit{denoising autoencoder} and \textit{equalization layers} --- we trained and tested the unmodified ResNet model, and as shown {in Table~\ref{tab:ASRperf}}, the performance drops by 10.37\% percent. 
Using a 2 m sampling tube reduces our model's performance to 72.8\% (see Appx.~\ref{appendix:futureWork}). 
Subsequently, we evaluate BaroVox in different scenarios that influence the recovered speech quality.

\subsubsection{\textbf{Performance analysis under various scenarios}}

\textbf{Evaluation setup}. 
We investigate BaroVox's responses to speaker volume, distance, and orientation variations.
\revision{Our experiments cover distances from 5cm to 2m and sound levels from 65dB to 90dB, reflecting common deployments in cleanrooms and other DPS environments. This range of configurations allows us to evaluate the attack's viability across realistic scenarios.}
Initially, the model’s generalized parameters led to suboptimal performance due to their lack of specificity to the nuanced conditions of each scenario. 
Without fine-tuning, the model struggles with reduced SNRs during lower volume levels or increased distances, impairing its classification accuracy. Different orientations present additional challenges by introducing phase variations and amplitude alterations, which the unadjusted model could not effectively handle.
To mitigate these performance issues, fine-tuning is deemed essential. The process involves retraining our ASR model using scenario-specific datasets, each reflecting the unique acoustical and signal characteristics associated with particular volume, distance, and orientation variations. Through this focused retraining, the model’s parameters are recalibrated, allowing it to more accurately navigate and adapt to the challenges within each scenario, leading to improved performance.

\textbf{Varying the volume of the audio source}. To investigate the speaker volume’s effect (measured in dB) on audio recovery, we record pressure readings at {{$90~dB$, $80~dB$, and $65~dB$}} of volume. 
Table~\ref{tab:factorsASR} reveals a significant correlation between volume and the classification model’s efficacy. 
As the volume decreases from $90~dB$ to $65~dB$, there's a notable decline in accuracy from {90.51\% to 81.38\%}. Furthermore, Table~\ref{tab:factorsASR} shows fine-tuning the model for this specific scenario boosted ASR model performance by a huge margin.

\begin{table}[h!]
    
		\footnotesize
  \centering    
    \begin{tabular}{|P{0.9cm}|P{0.98cm}|P{1.5cm}|P{1.5cm}|P{1.5cm}|}
    \hline
 \rowcolor{lightgray} \multicolumn{2}{|c|}{\multirow{3}{*}}  & \multicolumn{3}{c|}{\textbf{Performance} (in \%)} \\ \cline{3-5}
\rowcolor{lightgray}\multicolumn{2}{|c|}{ \textbf{Factors}} &  \multicolumn{3}{c|}{\textbf{\finetunned{Fine-tunned} / Unmodified}} \\ \cline{3-5}
\rowcolor{lightgray} \multicolumn{2}{|c|}{}&  \multicolumn{3}{c|} {\textbf{Speaker Orientation}} \\ \cline{3-5}
         \hline
         \rowcolor{lightgray}  \textbf{Volume}& \textbf{Distance}&$0$\textdegree& $90$\textdegree& $180$\textdegree\\
         \hline

         \multirow{3}{*}{\textbf{$90 dB$}} & $5 cm$& \finetunned{90.51}    / 90.51  & \finetunned{82.16} / 13.87  & \finetunned{82.19} / 10.15 \\ \cline{2-5}
                         &  $50 cm$&  \finetunned{78.75} / 7.24  & \finetunned{55.00} / 4.27 & \finetunned{69.81} /  7.24\\ \cline{2-5}   
                         &  $1 m$&\finetunned{78.27} / 2.48 &\finetunned{23.27} /  2.57 & \finetunned{30.76} /  3.41  \\\hline
         \multirow{3}{*}{}& $5 cm$ &\finetunned{86.65} / 12.91 & \finetunned{86.47} / 13.61   & \finetunned{57.45} / 9.47 \\ \cline{2-5}
         \textbf{$80 dB$} &  $50 cm$ & \finetunned{46.21} / 6.45  & \finetunned{43.50} / 3.82  & \finetunned{31.13} / 6.62 \\ \cline{2-5}
          &  $1 m$&  \finetunned{68.59} / 2.42 & \finetunned{19.94} / 1.97  & \finetunned{49.25} / 2.58 \\ \hline
        & $5cm$ &  \finetunned{81.38} / 17.13  & \finetunned{75.44} / 12.39   & \finetunned{80.83} / 8.98 \\ \cline{2-5}
        \textbf{$65 dB$} &  $50cm$ & \finetunned{65.68 } / 6.19 & \finetunned{77.36} / 3.46  & \finetunned{55.34} / 6.03  \\ \cline{2-5}
         &  $1m$ &\finetunned{56.30} / 2.34  & \finetunned{25.80} /  2.43 & \finetunned{33.71} / 3.00 \\ \hline
    \end{tabular}\caption{\label{tab:factorsASR}Performance of BaroVox in different scenarios. }
    \vspace{-1.5em}
\end{table}

{\textbf{Varying the distance of the audio source}}.
\label{subsection:distanceeffect} 
We explore the effect of increasing speaker-DPS distance on attack accuracy, testing at distances of $5cm$, $50cm$, and $1m$. 
The results are depicted in Table~\ref{tab:factorsASR}. Unsurprisingly, the model's performance varies inversely to the distance, given that sound amplitudes drop off with respect to the square of the distance. 
\revision{While the attack's effectiveness decreases with distance, this is a fundamental limitation shared by all sensor-based side-channel attacks that rely on sound waves. However, our attack model demonstrates robust performance up to 2m, which is comparable to or exceeds the effective range of many other sensor-based side-channel attacks. This range is sufficient for numerous real-world scenarios, particularly in cleanrooms and healthcare settings where DPS and sound sources are often in close proximity.
}

To further evaluate the BaroVox's robustness at a larger distance, we experimented by putting the speaker at a distance of  {$2m$} from the DPS. The classifiers' accuracy was 36.09\%,  much higher than random-guess accuracy. 
Nevertheless, the attacker must employ improved ASR models to overcome the distance limitation. 
We defer this task to future research, but the motivation is explained in Appx.~\ref{appendix:futureWork}.

{\textbf{Varying the orientation of the sound source from the DSP}}.
Given that human speech and speakers produce directed sound waves, we probe how a source's angle to the sensor influences ASR performance. We mount the pressure sensor at 0\textdegree,  90\textdegree, and 180\textdegree \,(when the speaker and DPS face similar direction) from the speaker and we report the result in Table~\ref{tab:factorsASR}. 
The results indicate that there exists a considerable effect on accuracy. The accuracy is higher at 0\textdegree \,because the sound wave's energy will be focused in the direction of the DPS, potentially creating a strong vibration.  The performance reduced to {82.19\% accuracy at 180\textdegree}.

\section{Discussion}

\subsection{Potential Outcomes and their Implications}
Our study reveals a critical security vulnerability in DPSs, demonstrating the feasibility of extracting speech from pressure data. 

\subsubsection{Information leakage across industries}
Our evaluation demonstrates significant potential for information leakage.
The Mean Opinion Score (MOS) of 3.36/5 for semiconductor-focused sentences indicates substantial semantic content in the reconstructed speech. 
The semiconductor industry exemplifies the potential impact of BaroVox. Given past IP theft incidents and the presence of sensitive information in cleanrooms (Appendices~\ref{subsec:CleanroomSecurity and IPs} and \ref{appendix:semiconductorInfo}), even fragments of reconstructed information could be highly valuable to competitors. In healthcare settings, similar reconstruction rates could lead to breaches of patient confidentiality, potentially violating regulations and trust.

\subsubsection{Automated threat scaling}
Our ASR model's 90.51\% accuracy on the SpeechCommand256 dataset represents a significant threat escalation. This high accuracy enables potential automated, large-scale eavesdropping in real-world scenarios. It could facilitate continuous monitoring, data mining of partially reconstructed speech, and contextual attacks based on identified key terms.

The combination of accurate speech reconstruction, high-performance ASR, and increasing connectivity of DPS in smart building systems creates a scenario where automated eavesdropping becomes a tangible threat. 

\subsubsection{Attack effectiveness in various conditions}
Our comprehensive evaluation of BaroVox reveals its effectiveness across various real-world conditions. The attack demonstrates high accuracy at close range (90.51\% at 5cm) and maintains significant effectiveness up to 1m (78.27\% after fine-tuning), with detectable speech components persisting at 2m. This performance curve aligns with many real-world DPS deployments. Notably, BaroVox adapts well to different volume levels, maintaining 81.38\% accuracy at conversational volume (65dB) at close range. The attack also shows remarkable resilience to source orientation, with accuracy remaining above 82\% even at perpendicular and opposite angles to the sound source.

These results indicate BaroVox's adaptability to diverse environments, from industrial settings to quieter spaces like offices or healthcare facilities. When compared to other sensor-based side-channel attacks, BaroVox demonstrates comparable or superior performance, suggesting that DPS are as vulnerable to acoustic side-channel attacks as sensors more commonly associated with such vulnerabilities. The significant improvements achieved through fine-tuning (e.g., from 2.48\% to 78.27\% at 1m) indicate potential for further enhancements, implying that the attack's effectiveness could increase with more sophisticated processing techniques. In conclusion, BaroVox presents a practical and adaptable attack vector, with its performance characteristics closely aligning with real-world DPS deployment scenarios, underscoring the urgent need for comprehensive countermeasures.

\subsubsection{Technological trends and future enhancements}
The increasing integration of DPS into IoT and smart building systems amplifies BaroVox's potential impact. The trend towards networked, often under-secured sensors significantly expands the attack surface. Future enhancements could further increase the threat level, including advanced signal processing techniques like adaptive noise cancellation, machine learning improvements such as attention mechanisms or transformers, and multi-sensor fusion in environments with multiple DPS. These improvements could extend the attack's effective range beyond 2m and enhance its performance in challenging acoustic environments.

\color{black}
\subsection{Limitation}
\revision{
While our research demonstrates the significant potential of BaroVox, it's important to consider the context and limitations of our findings. The attack's effectiveness depends on the attacker's ability to access pressure sensor readings, which varies across different deployment scenarios. This aspect highlights the importance of secure data handling practices in DPS-equipped environments.
}

\revision{
The performance of BaroVox is influenced by the specific characteristics of the DPS employed and environmental factors such as the proximity of sound sources. Our experiments with the ASR model in DS-II showed that fine-tuning might be necessary to adapt to diverse scenarios, indicating the attack's adaptability but also the need for scenario-specific optimizations.
}

\revision{
Our controlled experiments, while providing crucial insights, represent a first step in understanding this vulnerability. Real-world environments may present additional complexities not fully captured in our current study. 
Nonetheless, our work serves as a foundation for understanding the risks associated with the acoustic side-channel vulnerability in DPS and highlights the need for further research and development of countermeasures.
}

\subsection{Countermeasures}
Potential defenses against BaroVox attacks include:

{\textbf{Sound dampening}}. Dampening the sound wave by putting a sound-dampening material around the pressure ports is an inexpensive countermeasure. We conduct sound-dampening experiments using 3 materials: acrylic sheet, foam, and paper box. Using these materials, the performance of BaroVox decreased to $4.13\%$, $3.95\%$, and $4.04\%$, respectively.

{\textbf{Filtering}}. Incorporating a low-pass filter into the electronic components of the DPS can help mitigate the attack. The low-pass filter smooths the pressure readings and removes higher-frequency data representing speech, making it more difficult for an attacker to extract sensitive information. In our experiments, a third-order Butterworth low-pass filter with a cutoff frequency of 40 Hz successfully prevented the attack.

{\textbf{Increasing audio source distance}}. The proximity of the audio source to the DPS impacts the attack's efficacy. 
Our research indicates that placing the DPS at a distance exceeding $3.5$ m from sound sources effectively thwarts the attack.

\section{Conclusion}
We introduce BaroVox, a novel side-channel attack that exploits the acoustic vulnerabilities of DPS to reconstruct speech from pressure readings. Our two design solutions, focusing on signal processing and deep learning, demonstrate the effectiveness of BaroVox in recovering sensitive information. The implications of this attack extend beyond information leakage, potentially impacting finances, competitiveness, and security. Our work highlights the need for increased awareness and development of countermeasures to mitigate the risks posed by BaroVox. 

\section*{Acknowledgment}

The authors would like to thank our shepherd and the anonymous reviewers for their valuable comments, which greatly improved this paper. We also express our gratitude to Harsh Thomare for his contributions during the early stages of this research. 
This work was partially supported by the National Science Foundation (NSF) under award ECCS-2028269. Any opinions, findings, conclusions, or recommendations expressed in this paper are those of the authors and do not necessarily reflect the views of the funding agencies.


\bibliographystyle{IEEEtran}
\bibliography{barovox}

\begin{thebibliography}{10}
\providecommand{\url}[1]{#1}
\csname url@samestyle\endcsname
\providecommand{\newblock}{\relax}
\providecommand{\bibinfo}[2]{#2}
\providecommand{\BIBentrySTDinterwordspacing}{\spaceskip=0pt\relax}
\providecommand{\BIBentryALTinterwordstretchfactor}{4}
\providecommand{\BIBentryALTinterwordspacing}{\spaceskip=\fontdimen2\font plus
\BIBentryALTinterwordstretchfactor\fontdimen3\font minus \fontdimen4\font\relax}
\providecommand{\BIBforeignlanguage}[2]{{%
\expandafter\ifx\csname l@#1\endcsname\relax
\typeout{** WARNING: IEEEtran.bst: No hyphenation pattern has been}%
\typeout{** loaded for the language `#1'. Using the pattern for}%
\typeout{** the default language instead.}%
\else
\language=\csname l@#1\endcsname
\fi
#2}}
\providecommand{\BIBdecl}{\relax}
\BIBdecl

\bibitem{avnet2024pressure}
\BIBentryALTinterwordspacing
A.~Abacus, ``Pressure sensors: The design engineer's guide,'' 2017, retrieved from Avnet website. [Online]. Available: \url{https://my.avnet.com/abacus/solutions/technologies/sensors/pressure-sensors/}
\BIBentrySTDinterwordspacing

\bibitem{standard2004cleanrooms}
C.~Standard, British and B.~ISO, ``Cleanrooms and associated controlled environments—,'' 2004.

\bibitem{datasheetsdp}
\BIBentryALTinterwordspacing
T.~S. Company, ``Datasheet sdp8xx-digital differential pressure sensor,'' 2019, \url. (Accessed: 05-21-2024). [Online]. Available: \url{{https://sensirion.com/media/documents/90500156/6167E43B/Sensirion_Differential_Pressure_Datasheet_SDP8xx_Digital.pdf}}
\BIBentrySTDinterwordspacing

\bibitem{bradford_hvac_2024}
\BIBentryALTinterwordspacing
C.~Bradford, ``Understanding your hvac system: Building pressure monitoring and control,'' Buildings IOT website, 2024, accessed: 05-21-2024. [Online]. Available: \url{https://www.buildingsiot.com/blog/understanding-your-hvac-system-building-pressure-monitoring-and-control-bd}
\BIBentrySTDinterwordspacing

\bibitem{fitzgerald2010harmonic}
D.~Fitzgerald, ``Harmonic/percussive separation using median filtering,'' in \emph{Proceedings of the International Conference on Digital Audio Effects (DAFx)}, vol.~13, 2010, pp. 1--4.

\bibitem{driedger2014extending}
J.~Driedger, M.~M{\"u}ller, and S.~Disch, ``Extending harmonic-percussive separation of audio signals.'' in \emph{ISMIR}, 2014, pp. 611--616.

\bibitem{leng2021mbnet}
Y.~Leng, X.~Tan, S.~Zhao, F.~Soong, X.-Y. Li, and T.~Qin, ``Mbnet: Mos prediction for synthesized speech with mean-bias network,'' in \emph{ICASSP 2021-2021 IEEE International Conference on Acoustics, Speech and Signal Processing (ICASSP)}.\hskip 1em plus 0.5em minus 0.4em\relax IEEE, 2021, pp. 391--395.

\bibitem{errattahi2018automatic}
R.~Errattahi, A.~El~Hannani, and H.~Ouahmane, ``Automatic speech recognition errors detection and correction: A review,'' \emph{Procedia Computer Science}, vol. 128, pp. 32--37, 2018.

\bibitem{firstsensor:2020}
\BIBentryALTinterwordspacing
F.~S. AG, ``How to decide on piezoresistive or thermal measuring principle,'' \emph{AZoSensors}, October 13 2020. [Online]. Available: \url{https://www.azosensors.com/article.aspx?ArticleID=1723}
\BIBentrySTDinterwordspacing

\bibitem{cooper1986particulate}
D.~W. Cooper, ``Particulate contamination and microelectronics manufacturing: an introduction,'' \emph{Aerosol Science and Technology}, vol.~5, no.~3, pp. 287--299, 1986.

\bibitem{kitajima1997requirements}
H.~Kitajima and Y.~Shiramizu, ``Requirements for contamination control in the gigabit era,'' \emph{IEEE transactions on semiconductor manufacturing}, vol.~10, no.~2, pp. 267--272, 1997.

\bibitem{shilov_2019}
\BIBentryALTinterwordspacing
A.~Shilov, ``{TSMC's Fab 14B Photoresist Material Incident: \$550 Million in Lost Revenue},'' \emph{AnandTech}, 2019. [Online]. Available: \url{https://www.anandtech.com/show/13975/tsmcs-fab-14b-photoresist-material-incident-550-million-in-lost-revenue}
\BIBentrySTDinterwordspacing

\bibitem{yap_2018}
\BIBentryALTinterwordspacing
L.~Yap, ``{Samsung's Production Plant Contaminated Resulting in \$560 Million Loss},'' \emph{Tech Critter}, 2018. [Online]. Available: \url{https://www.tech-critter.com/samsung-manufacturing-plant-contamination/}
\BIBentrySTDinterwordspacing

\bibitem{miller2017implementing}
S.~L. Miller, N.~Clements, S.~A. Elliott, S.~S. Subhash, A.~Eagan, and L.~J. Radonovich, ``Implementing a negative-pressure isolation ward for a surge in airborne infectious patients,'' \emph{American journal of infection control}, vol.~45, no.~6, pp. 652--659, 2017.

\bibitem{litheaudio2023}
\BIBentryALTinterwordspacing
L.~Audio, ``Healthcare operating theatre audio,'' 2023, accessed: 05-21-2024. [Online]. Available: \url{https://www.litheaudio.com/healthcare-operating-theatre-audio.html}
\BIBentrySTDinterwordspacing

\bibitem{zenitel2023}
\BIBentryALTinterwordspacing
Zenitel, ``Cleanroom intercom station ip-cror datasheet,'' 2023, accessed: 05-21-2024. [Online]. Available: \url{https://www.zenitel.com/print/pdf/node/4584}
\BIBentrySTDinterwordspacing

\bibitem{michalevsky2014gyrophone}
Y.~Michalevsky, D.~Boneh, and G.~Nakibly, ``Gyrophone: Recognizing speech from gyroscope signals,'' in \emph{23rd $\{$USENIX$\}$ Security Symposium ($\{$USENIX$\}$ Security 14)}, 2014, pp. 1053--1067.

\bibitem{zhang2015accelword}
L.~Zhang, P.~H. Pathak, M.~Wu, Y.~Zhao, and P.~Mohapatra, ``Accelword: Energy efficient hotword detection through accelerometer,'' in \emph{Proceedings of the 13th Annual International Conference on Mobile Systems, Applications, and Services}, 2015, pp. 301--315.

\bibitem{anand2018speechless}
S.~A. Anand and N.~Saxena, ``Speechless: Analyzing the threat to speech privacy from smartphone motion sensors,'' in \emph{2018 IEEE Symposium on Security and Privacy (SP)}.\hskip 1em plus 0.5em minus 0.4em\relax IEEE, 2018, pp. 1000--1017.

\bibitem{anand2021spearphone}
S.~A. Anand, C.~Wang, J.~Liu, N.~Saxena, and Y.~Chen, ``Spearphone: a lightweight speech privacy exploit via accelerometer-sensed reverberations from smartphone loudspeakers,'' in \emph{Proceedings of the 14th ACM Conference on Security and Privacy in Wireless and Mobile Networks}, 2021, pp. 288--299.

\bibitem{han2017pitchln}
J.~Han, A.~J. Chung, and P.~Tague, ``Pitchln: eavesdropping via intelligible speech reconstruction using non-acoustic sensor fusion,'' in \emph{Proceedings of the 16th ACM/IEEE International Conference on Information Processing in Sensor Networks}, 2017, pp. 181--192.

\bibitem{7878599}
S.~Rokka~Chhetri and M.~A. Al~Faruque, ``Side channels of cyber-physical systems: Case study in additive manufacturing,'' \emph{IEEE Design \& Test}, vol.~34, no.~4, pp. 18--25, 2017.

\bibitem{marquardt2011sp}
P.~Marquardt, A.~Verma, H.~Carter, and P.~Traynor, ``(sp) iphone: Decoding vibrations from nearby keyboards using mobile phone accelerometers,'' in \emph{Proceedings of the 18th ACM conference on Computer and communications security}, 2011, pp. 551--562.

\bibitem{9517289}
Y.~Zhang, R.~Yasaei, H.~Chen, Z.~Li, and M.~A. Al~Faruque, ``Stealing neural network structure through remote fpga side-channel analysis,'' \emph{IEEE Transactions on Information Forensics and Security}, vol.~16, pp. 4377--4388, 2021.

\bibitem{9095984}
M.~A. Al~Faruque, S.~R. Chhetri, A.~Canedo, and J.~Wan, ``Acoustic side-channel attacks on additive manufacturing systems,'' in \emph{2016 ACM/IEEE 7th International Conference on Cyber-Physical Systems (ICCPS)}, 2016, pp. 1--10.

\bibitem{cpams}
\BIBentryALTinterwordspacing
S.~R. Chhetri, A.~Canedo, and M.~A.~A. Faruque, ``Confidentiality breach through acoustic side-channel in cyber-physical additive manufacturing systems,'' \emph{ACM Trans. Cyber-Phys. Syst.}, vol.~2, no.~1, Dec. 2017. [Online]. Available: \url{https://doi.org/10.1145/3078622}
\BIBentrySTDinterwordspacing

\bibitem{ba2020learning}
Z.~Ba, T.~Zheng, X.~Zhang, Z.~Qin, B.~Li, X.~Liu, and K.~Ren, ``Learning-based practical smartphone eavesdropping with built-in accelerometer.'' in \emph{NDSS}, 2020, pp. 23--26.

\bibitem{wang2022mmphone}
C.~Wang, F.~Lin, T.~Liu, Z.~Liu, Y.~Shen, Z.~Ba, L.~Lu, W.~Xu, and K.~Ren, ``mmphone: Acoustic eavesdropping on loudspeakers via mmwave-characterized piezoelectric effect,'' in \emph{IEEE INFOCOM 2022-IEEE Conference on Computer Communications}.\hskip 1em plus 0.5em minus 0.4em\relax IEEE, 2022, pp. 820--829.

\bibitem{gao2022device}
M.~Gao, Y.~Liu, Y.~Chen, Y.~Li, Z.~Ba, X.~Xu, J.~Han, and K.~Ren, ``Device-independent smartphone eavesdropping jointly using accelerometer and gyroscope,'' \emph{IEEE Transactions on Dependable and Secure Computing}, 2022.

\bibitem{gao2022kite}
M.~Gao, L.~Zhang, L.~Shen, X.~Zou, J.~Han, F.~Lin, and K.~Ren, ``Kite: exploring the practical threat from acoustic transduction attacks on inertial sensors,'' in \emph{Proceedings of the 20th ACM Conference on Embedded Networked Sensor Systems}, 2022, pp. 696--709.

\bibitem{sami2020spying}
S.~Sami, Y.~Dai, S.~R.~X. Tan, N.~Roy, and J.~Han, ``Spying with your robot vacuum cleaner: eavesdropping via lidar sensors,'' in \emph{Proceedings of the 18th Conference on Embedded Networked Sensor Systems}, 2020, pp. 354--367.

\bibitem{roy2016listening}
N.~Roy and R.~Roy~Choudhury, ``Listening through a vibration motor,'' in \emph{Proceedings of the 14th Annual International Conference on Mobile Systems, Applications, and Services}, 2016, pp. 57--69.

\bibitem{nassi2022lamphone}
B.~Nassi, Y.~Pirutin, R.~Swisa, A.~Shamir, Y.~Elovici, and B.~Zadov, ``Lamphone: Passive sound recovery from a desk lamp's light bulb vibrations,'' in \emph{31st USENIX Security Symposium (USENIX Security 22)}, 2022, pp. 4401--4417.

\bibitem{kwong2019hard}
A.~Kwong, W.~Xu, and K.~Fu, ``Hard drive of hearing: Disks that eavesdrop with a synthesized microphone,'' in \emph{2019 IEEE symposium on security and privacy (SP)}.\hskip 1em plus 0.5em minus 0.4em\relax IEEE, 2019, pp. 905--919.

\bibitem{long2023side}
Y.~Long, P.~Naghavi, B.~Kojusner, K.~Butler, S.~Rampazzi, and K.~Fu, ``Side eye: Characterizing the limits of pov acoustic eavesdropping from smartphone cameras with rolling shutters and movable lenses,'' \emph{arXiv preprint arXiv:2301.10056}, 2023.

\bibitem{nassi2021glowworm}
B.~Nassi, Y.~Pirutin, T.~Galor, Y.~Elovici, and B.~Zadov, ``Glowworm attack: Optical tempest sound recovery via a device's power indicator led,'' in \emph{Proceedings of the 2021 ACM SIGSAC Conference on Computer and Communications Security}, 2021, pp. 1900--1914.

\bibitem{nassi2023little}
B.~Nassi, R.~Swissa, J.~Shams, B.~Zadov, and Y.~Elovici, ``The little seal bug: Optical sound recovery from lightweight reflective objects,'' in \emph{2023 IEEE Security and Privacy Workshops (SPW)}.\hskip 1em plus 0.5em minus 0.4em\relax IEEE, 2023, pp. 298--310.

\bibitem{nassi2022little}
B.~Nassi, R.~Swissa, Y.~Elovici, and B.~Zadov, ``The little seal bug: Optical sound recovery from lightweight reflective objects.'' \emph{IACR Cryptol. ePrint Arch.}, vol. 2022, p. 227, 2022.

\bibitem{wang2022wavesdropper}
C.~Wang, F.~Lin, Z.~Ba, F.~Zhang, W.~Xu, and K.~Ren, ``Wavesdropper: Through-wall word detection of human speech via commercial mmwave devices,'' \emph{Proceedings of the ACM on Interactive, Mobile, Wearable and Ubiquitous Technologies}, vol.~6, no.~2, pp. 1--26, 2022.

\bibitem{basak2022mmspy}
S.~Basak and M.~Gowda, ``mmspy: Spying phone calls using mmwave radars,'' in \emph{2022 IEEE Symposium on Security and Privacy (SP)}.\hskip 1em plus 0.5em minus 0.4em\relax IEEE, 2022, pp. 1211--1228.

\bibitem{tu2021transduction}
Y.~Tu, V.~S. Tida, Z.~Pan, and X.~Hei, ``Transduction shield: A low-complexity method to detect and correct the effects of emi injection attacks on sensors,'' in \emph{Proceedings of the 2021 ACM Asia Conference on Computer and Communications Security}, 2021, pp. 901--915.

\bibitem{barua2022wolf}
A.~Barua, Y.~G. Achamyeleh, and M.~A. Al~Faruque, ``A wolf in sheep's clothing: Spreading deadly pathogens under the disguise of popular music,'' in \emph{Proceedings of the 2022 ACM SIGSAC Conference on Computer and Communications Security}, 2022, pp. 277--291.

\bibitem{primex2021}
Primex, ``Environmental monitoring solutions: Room pressure monitoring,'' \url{https://www.primexinc.com/en/solutions/environmental-monitoring/room-pressure-monitoring}, 2021, accessed: 05-21-2024.

\bibitem{lee2019securely}
S.~Lee, W.~Choi, H.~J. Jo, and D.~H. Lee, ``How to securely record logs based on arm trustzone,'' in \emph{Proceedings of the 2019 ACM Asia Conference on Computer and Communications Security}, 2019, pp. 664--666.

\bibitem{paccagnella2020logging}
R.~Paccagnella, K.~Liao, D.~Tian, and A.~Bates, ``Logging to the danger zone: Race condition attacks and defenses on system audit frameworks,'' in \emph{Proceedings of the 2020 ACM SIGSAC Conference on Computer and Communications Security}, 2020, pp. 1551--1574.

\bibitem{xb13}
\BIBentryALTinterwordspacing
Sony, ``Xb13 extra bass™ portable wireless speaker,'' 2021. [Online]. Available: \url{https://www.sony.com/electronics/support/res/manuals/5025/f505fde429679d4719abd78c1a231ac4/50254675M.pdf}
\BIBentrySTDinterwordspacing

\bibitem{rs-online}
R.~Group, ``{Datasheet: Raspberry Pi 3 Model B},'' \url{https://us.rs-online.com/m/d/4252b1ecd92888dbb9d8a39b536e7bf2.pdf}, 2018, accessed: 05-21-2024.

\bibitem{vaidyanathan2001generalizations}
P.~Vaidyanathan, ``Generalizations of the sampling theorem: Seven decades after nyquist,'' \emph{IEEE Transactions on Circuits and Systems I: Fundamental Theory and Applications}, vol.~48, no.~9, pp. 1094--1109, 2001.

\bibitem{heide1998speech}
D.~A. Heide and G.~S. Kang, ``Speech enhancement for bandlimited speech,'' in \emph{Proceedings of the 1998 IEEE International Conference on Acoustics, Speech and Signal Processing, ICASSP'98 (Cat. No. 98CH36181)}, vol.~1.\hskip 1em plus 0.5em minus 0.4em\relax IEEE, 1998, pp. 393--396.

\bibitem{villchur1973signal}
E.~Villchur, ``Signal processing to improve speech intelligibility in perceptive deafness,'' \emph{The Journal of the Acoustical Society of America}, vol.~53, no.~6, pp. 1646--1657, 1973.

\bibitem{whitmore2009improved}
S.~A. Whitmore and B.~Fox, ``Improved accuracy, second-order response model for pressure sensing systems,'' \emph{Journal of aircraft}, vol.~46, no.~2, pp. 491--500, 2009.

\bibitem{ladefoged2014course}
P.~Ladefoged and K.~Johnson, \emph{A course in phonetics}.\hskip 1em plus 0.5em minus 0.4em\relax Cengage learning, 2014.

\bibitem{gilman2019science}
M.~Gilman, ``The science of voice and the body,'' \emph{The Oxford handbook of music and the body}, pp. 62--78, 2019.

\bibitem{moulines1995non}
E.~Moulines and J.~Laroche, ``Non-parametric techniques for pitch-scale and time-scale modification of speech,'' \emph{Speech communication}, vol.~16, no.~2, pp. 175--205, 1995.

\bibitem{cano2014phase}
E.~Cano, M.~Plumbley, and C.~Dittmar, ``Phase-based harmonic/percussive separation,'' in \emph{Fifteenth Annual Conference of the International Speech Communication Association}, 2014.

\bibitem{tachibana2014harmonic}
H.~Tachibana, N.~Ono, H.~Kameoka, and S.~Sagayama, ``Harmonic/percussive sound separation based on anisotropic smoothness of spectrograms,'' \emph{IEEE/ACM Transactions on Audio, Speech, and Language Processing}, vol.~22, no.~12, pp. 2059--2073, 2014.

\bibitem{zhang2021influence}
L.~Zhang, F.~Schlaghecken, J.~Harte, and K.~L. Roberts, ``The influence of the type of background noise on perceptual learning of speech in noise,'' \emph{Frontiers in Neuroscience}, vol.~15, p. 646137, 2021.

\bibitem{nishimura2004noise}
Y.~Nishimura, T.~Shinozaki, K.~Iwano, and S.~Furui, ``Noise-robust speech recognition using multi-band spectral features,'' \emph{The Journal of the Acoustical Society of America}, vol. 116, no.~4, pp. 2480--2480, 2004.

\bibitem{li2020time}
S.~Li, W.~Yuan, J.~Yuan, B.~Bai, D.~W.~K. Ng, and L.~Hanzo, ``Time-domain vs. frequency-domain equalization for ftn signaling,'' \emph{IEEE transactions on vehicular technology}, vol.~69, no.~8, pp. 9174--9179, 2020.

\bibitem{falconer2002frequency}
D.~Falconer, S.~L. Ariyavisitakul, A.~Benyamin-Seeyar, and B.~Eidson, ``Frequency domain equalization for single-carrier broadband wireless systems,'' \emph{IEEE Communications Magazine}, vol.~40, no.~4, pp. 58--66, 2002.

\bibitem{speechcommandsv2}
\BIBentryALTinterwordspacing
P.~{Warden}, ``{Speech Commands: A Dataset for Limited-Vocabulary Speech Recognition},'' \emph{ArXiv e-prints}, Apr. 2018. [Online]. Available: \url{https://arxiv.org/abs/1804.03209}
\BIBentrySTDinterwordspacing

\bibitem{ts1}
B.~Zhao, H.~Lu, S.~Chen, J.~Liu, and D.~Wu, ``Convolutional neural networks for time series classification,'' \emph{Journal of Systems Engineering and Electronics}, vol.~28, no.~1, pp. 162--169, 2017.

\bibitem{ts2}
F.~Karim, S.~Majumdar, H.~Darabi, and S.~Chen, ``Lstm fully convolutional networks for time series classification,'' \emph{IEEE Access}, vol.~6, pp. 1662--1669, 2018.

\bibitem{ts3}
Z.~Cui, W.~Chen, and Y.~Chen, ``Multi-scale convolutional neural networks for time series classification,'' 2016.

\bibitem{resnet}
K.~He, X.~Zhang, S.~Ren, and J.~Sun, ``Deep residual learning for image recognition,'' 2015.

\bibitem{audioresnet}
S.~Hershey, S.~Chaudhuri, D.~P. Ellis, J.~F. Gemmeke, A.~Jansen, R.~C. Moore, M.~Plakal, D.~Platt, R.~A. Saurous, B.~Seybold \emph{et~al.}, ``Cnn architectures for large-scale audio classification,'' in \emph{2017 ieee international conference on acoustics, speech and signal processing (icassp)}.\hskip 1em plus 0.5em minus 0.4em\relax IEEE, 2017, pp. 131--135.

\bibitem{receptivefields}
K.~Koutini, H.~Eghbal-zadeh, and G.~Widmer, ``Receptive field regularization techniques for audio classification and tagging with deep convolutional neural networks,'' \emph{IEEE/ACM Transactions on Audio, Speech, and Language Processing}, vol.~29, pp. 1987--2000, 2021.

\bibitem{gu2021efficiently}
A.~Gu, K.~Goel, and C.~R{\'e}, ``Efficiently modeling long sequences with structured state spaces,'' \emph{arXiv preprint arXiv:2111.00396}, 2021.

\bibitem{cobo2008airborne}
F.~Cobo, D.~Grela, and A.~n. Conchal, ``Airborne particle monitoring in clean room environments for stem cell cultures,'' \emph{Biotechnology Journal: Healthcare Nutrition Technology}, vol.~3, no.~1, pp. 43--52, 2008.

\bibitem{holbrook2009controlling}
D.~Holbrook, ``Controlling contamination: the origins of clean room technology,'' \emph{History and Technology}, vol.~25, no.~3, pp. 173--191, 2009.

\bibitem{taipeitimes_tsmc_2017}
\BIBentryALTinterwordspacing
T.~Times, ``{TSMC} engineer charged with stealing trade secrets,'' \emph{Taipei Times}. [Online]. Available: \url{https://www.taipeitimes.com/News/biz/archives/2017/05/03/2003669834}
\BIBentrySTDinterwordspacing

\bibitem{justice_chinese_2018}
\BIBentryALTinterwordspacing
{U.S. Attorney's Office Northern District of California}, ``Chinese citizen sentenced for economic espionage, theft of trade secrets, and conspiracy,'' \emph{Department of Justice}. [Online]. Available: \url{https://www.justice.gov/usao-ndca/pr/chinese-citizen-sentenced-economic-espionage-theft-trade-secrets-and-conspiracy}
\BIBentrySTDinterwordspacing

\bibitem{justice_chinese_2021}
\BIBentryALTinterwordspacing
{U.S. Attorney's Office District of Connecticut}, ``Three {Chinese} nationals arrested in scheme to steal and illegally export military-grade carbon fiber from the {U.S.}'' \emph{Department of Justice}. [Online]. Available: \url{https://www.justice.gov/usao-ct/pr/three-chinese-nationals-arrested-scheme-steal-and-illegally-export-military-grade}
\BIBentrySTDinterwordspacing

\bibitem{justice2018lexington}
{U.S. Department of Justice}, ``Lexington man and semiconductor company indicted for theft of trade secrets,'' \url{https://shorturl.at/imqyG}, 2018, accessed: 05-21-2024.

\bibitem{Alpha161}
\BIBentryALTinterwordspacing
A.~Instruments, ``{Alpha 161 Low-Cost Differential Pressure Transducer datasheet},'' 2021, accessed: 05-21-2024. [Online]. Available: \url{https://www.alphainstruments.com/product/model-161-differential-pressure-transmitter/}
\BIBentrySTDinterwordspacing

\bibitem{testo-6383}
\BIBentryALTinterwordspacing
T.~S.~. Co. (2017) Testo 6383 data sheet. Accessed: 05-21-2024. [Online]. Available: \url{https://static-int.testo.com/media/97/f5/9593ea116ffe/testo-6383-EN.pdf}
\BIBentrySTDinterwordspacing

\bibitem{siemens547}
\BIBentryALTinterwordspacing
I.~Siemens~Industry, ``Room pressure monitor, technical specification sheet,'' 2020, accessed: 05-21-2024. [Online]. Available: \url{https://sid.siemens.com/v/u/A6V10322677}
\BIBentrySTDinterwordspacing

\bibitem{PASCAL-ST}
\BIBentryALTinterwordspacing
N.~AG, ``{PASCAL-ST/ZB Accurate \& long-term stable measurement},'' 2016, accessed: 05-21-2024. [Online]. Available: \url{https://cesstech.com/wp-content/uploads/2023/07/Product-flyer_PascalST_ZB_EN_005391_00-1.pdf}
\BIBentrySTDinterwordspacing

\bibitem{RSMEB003}
\BIBentryALTinterwordspacing
I.~Dwyer~Instruments, ``Room status monitor,'' 2021, accessed: 05-21-2024. [Online]. Available: \url{https://www.dwyer-inst.com/PDF_files/RSME.pdf}
\BIBentrySTDinterwordspacing

\bibitem{sensos}
{SensoScientific, Inc.}, ``Differential pressure sensor data sheet,'' \url{https://www.laboratory-equipment.com/media/asset-library/d/i/differential-pressure-sensor-sensoscientific-data-sheet.pdf}, 2017, accessed: 05-21-2024.

\bibitem{spectrogramnfft}
\BIBentryALTinterwordspacing
E.~C. Knight, S.~P. Hernandez, E.~M. Bayne, V.~Bulitko, and B.~V. Tucker, ``Pre-processing spectrogram parameters improve the accuracy of bioacoustic classification using convolutional neural networks,'' \emph{Bioacoustics}, vol.~29, no.~3, pp. 337--355, 2020. [Online]. Available: \url{https://doi.org/10.1080/09524622.2019.1606734}
\BIBentrySTDinterwordspacing

\bibitem{subspectral}
S.~Chang, H.~Park, J.~Cho, H.~Park, S.~Yun, and K.~Hwang, ``Subspectral normalization for neural audio data processing,'' 2021.

\bibitem{kamal2022}
\BIBentryALTinterwordspacing
K.~Y. Kamal, ``The silicon age: Trends in semiconductor devices industry,'' \emph{Journal of Engineering Science and Technology Review}, 2022, accessed: 05-21-2024. [Online]. Available: \url{https://www.researchgate.net/publication/360851950_The_Silicon_Age_Trends_in_Semiconductor_Devices_Industry}
\BIBentrySTDinterwordspacing

\bibitem{nader2011harmonic}
C.~Nader, W.~Van~Moer, K.~Barbe, N.~Bjorsell, and P.~Handel, ``Harmonic sampling and reconstruction of wideband undersampled waveforms: Breaking the code,'' \emph{IEEE transactions on microwave theory and techniques}, vol.~59, no.~11, pp. 2961--2969, 2011.

\bibitem{nader2011unfolding}
C.~Nader, N.~Bj{\"o}rsell, and P.~H{\"a}ndel, ``Unfolding the frequency spectrum for undersampled wideband data,'' \emph{Signal Processing}, vol.~91, no.~5, pp. 1347--1350, 2011.

\bibitem{chung2021w2vbert}
Y.-A. Chung, Y.~Zhang, W.~Han, C.-C. Chiu, J.~Qin, R.~Pang, and Y.~Wu, ``W2v-bert: Combining contrastive learning and masked language modeling for self-supervised speech pre-training,'' 2021.

\bibitem{zhang2022pushing}
Y.~Zhang, J.~Qin, D.~S. Park, W.~Han, C.-C. Chiu, R.~Pang, Q.~V. Le, and Y.~Wu, ``Pushing the limits of semi-supervised learning for automatic speech recognition,'' 2022.

\bibitem{hasani2022liquid}
R.~Hasani, M.~Lechner, T.-H. Wang, M.~Chahine, A.~Amini, and D.~Rus, ``Liquid structural state-space models,'' \emph{arXiv preprint arXiv:2209.12951}, 2022.

\end{thebibliography}
\appendices
\section{Basics of Cleanrooms}
\label{appendix:cleanroombasics}
To assess the impact of targeted eavesdropping on sensitive information, this section explains cleanroom and IP in the semiconductor manufacturing industry,  including the implication of side-channel eavesdropping in these types of secure environments. 

\subsection{Cleanroom in semiconductor industry}
\label{subsec:Cleanroom}
Cleanrooms are controlled environments designed to filter out pollutants, ensuring that airborne contaminants remain at acceptable low-level concentrations~\cite{cobo2008airborne, standard2004cleanrooms}. 
While employed across various sectors ~\cite{holbrook2009controlling}, cleanrooms are paramount in semiconductor manufacturing where even a single speck of dust can drastically compromise chip quality~\cite{cooper1986particulate,
kitajima1997requirements}.
Consequently, cleanrooms are integral to assuring semiconductor product integrity.
Emphasizing their importance, contamination mishaps at giants like Samsung and TSMC have previously led to staggering combined losses of more than $\$1$ billion~\cite{shilov_2019, yap_2018}

\subsection{Cleanroom and Intellectual Property (IP)}
\label{subsec:CleanroomSecurity and IPs}

Cleanrooms not only protect the semiconductor industry from contaminants but also play a vital role by protecting proprietary products from unauthorized access and tampering. 
With \textbf{\textit{IP and national security}} at stake, a lapse in cleanroom security can be costly, especially amid the rising IP war.
Numerous theft incidents underscore this; for instance, TSMC grappled with a major trade secret theft attempt~\cite{taipeitimes_tsmc_2017}, and arrests have been made regarding IP theft from US semiconductor firms~\cite{justice_chinese_2018,justice_chinese_2021,justice2018lexington}.
To counter these threats, rigorous security protocols are enforced.

\subsection{{Pressure sensors used in cleanrooms}}
\label{subsection:psincleanrooms}

{ 
DPSs are at the heart of maintaining a cleanroom's integrity for chip manufacturing.
These sensors are crucial in regulating the airflow and maintaining a specific pressure level within the cleanroom. 
Specifically, the cleanroom must be maintained at a higher static pressure than adjacent spaces to prevent contaminants from entering. 
To achieve this, a Room Pressure Monitoring (RPM) system with an integrated pressure sensor offers real-time differential pressure tracking between distinct points.
Table \ref{table:DPS in cleanroom} 
shows popular manufacturers' RPM systems used in semiconductor cleanrooms, all predominantly incorporating DPSs, providing evidence of DPSs' industry prevalence.
}

\begin{table}[h!]
	\footnotesize
	\centering
		
		\begin{tabular}{p{0.12cm}|p{2.3cm}|p{1.5cm}|p{1.40cm}}
			\hline
			 \cellcolor [gray]{0.85} \textbf{Sl.} & \cellcolor [gray]{0.85} \textbf{RPM} &  \cellcolor [gray]{0.85} \textbf{Manufacturer} & \cellcolor [gray]{0.85} \textbf{Type}   \\
			\hline
			\hline
			1 & Alpha 161 \cite{Alpha161} & Alpha Inst. & Differential  \\
			\hline
    		2 & Testo 6383 \cite{testo-6383}  & Testo  & Differential\\
			\hline
                3 & Siemens 547-203 \cite{siemens547} & Siemens & Differential \\
			\hline
		    4 & PASCAL-ST/ZB \cite{PASCAL-ST} & Novasina & Differential  \\
			\hline
			5 & RSME-B-003 \cite{RSMEB003}& Dwyer    & Differential  \\
			\hline
			6 & B20-200-OTA \cite{sensos}& Sensoscientific   & Differential \\
			\hline			
		\end{tabular}
  \caption{\label{table:DPS in cleanroom}Differential pressure sensors used in cleanrooms.}
		
\end{table}
\vspace{-0.40em}


\section{Miscellaneous}

\subsection{Modeling effects of sound on DPS}
\label{appendix:modelingsound}
Fig.~\ref{fig:soundpressure} shows the model of sound signal a s a pressure wave and its effect on a pressure reading.
\begin{figure}[h]
\centering
\includegraphics[width=0.45\textwidth,height=0.205\textheight]{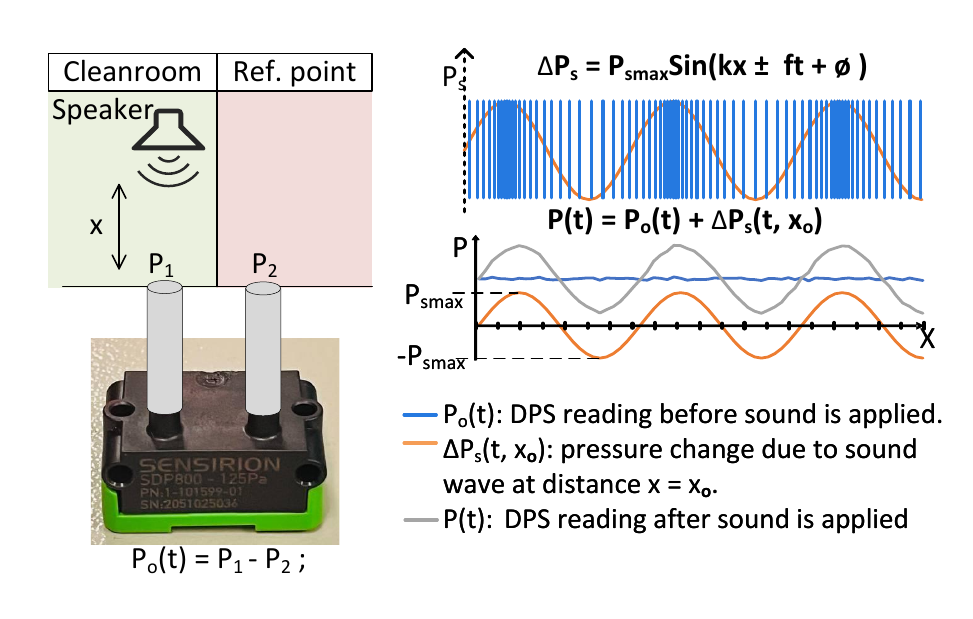}
\vspace{-1.00em}
\caption{Modeling sound signal as a pressure wave. }
\label{fig:soundpressure}
\vspace{-0.900em}
\end{figure}


\subsection{Percussive-Harmoinic Separation using Median Filtering}
\label{appendix:phs}
Fig~\ref{fig:harm_perc_filtering} visually illustrates the spectral subtraction process.
\begin{figure*}[h]
\centering
\includegraphics[trim={10px 59px 10px 25px},clip,width=0.85\linewidth]{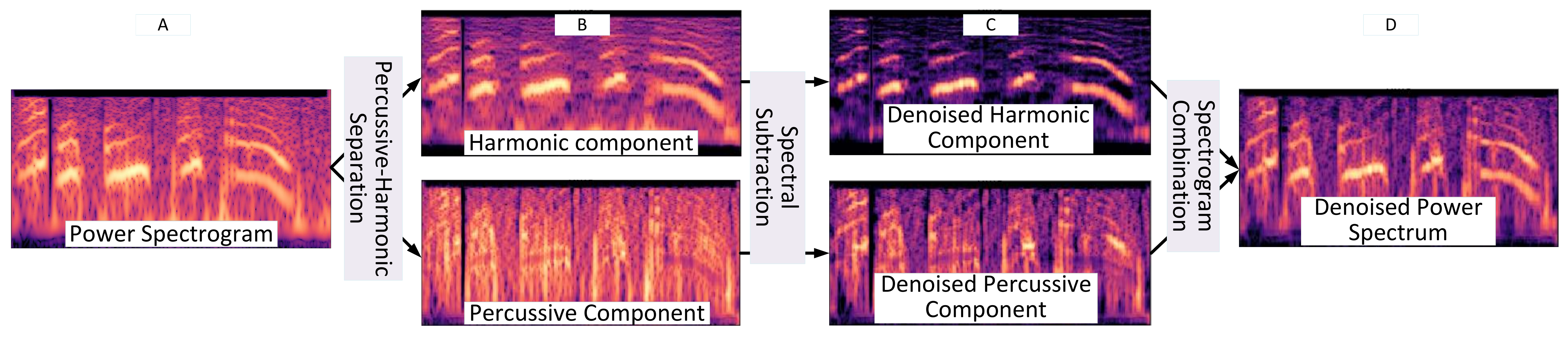}
\caption{Spectral subtraction by Percussive-Harmonic Separation using median filtering technique.}
\label{fig:harm_perc_filtering}
\end{figure*}

\subsection{SpeechCommands dataset composition}
\label{appendix:data}

The SpeechCommands v2 dataset ~\cite{speechcommandsv2} is a collection of spoken commands in English, consisting of $105,829$ utterances across 35 different words and phrases, such as "yes", "no", "stop", "go", "bed", "bird", "tree", and "wow". The dataset includes recordings from $2,618$ different speakers, spanning a wide range of ages and genders. The dataset was recorded in various acoustic conditions, including different background noise and reverberation levels, and contains both clean and noisy recordings. A complete list of the number of utterances per word can be found in ~\cite{speechcommandsv2}.

\section{ASR Model Architecture Summary}
\label{appendix:models}
\subsection{Spectrogram parameters}
This section provides an overview of the primary parameters of the $Spectrogram$ transform, which plays a vital role in time series signal processing. Understanding these parameters, including $NFFT$, window length, and window hop, is essential for effectively analyzing and interpreting time series data. For a more in-depth analysis of the effect of these parameters, we refer our readers to~\cite{spectrogramnfft}.

(1) \textbf{Number of FFT bins ($NFFT$)} is a parameter in the $Spectrogram$ process determining the frequency resolution of the spectrogram. A higher $NFFT$ value offers better frequency resolution but increased computational complexity. Alternatively, if the NFFT size is increased, the time resolution of the spectrogram will decrease. This occurs because the window used for each time segment becomes larger. As a result, rapid changes in the signal may be overlooked or smoothed out, leading to a loss of temporal information. It is important to tune the $NFFT$ value to fit the specific purpose of the application. Typically, $NFFT$ is set to a power of 2 for optimization. In our experiment $NFFT$ size of 256 results in better accuracy. 

(2) \textbf{Window length} refers to the length of the window function applied to audio segments in the $Spectrogram$ process. It affects both time and frequency resolutions, with longer windows providing better frequency resolution at the expense of time resolution. Typically, the window length is equal to or larger than $NFFT$.
    
(3) \textbf{Window hop} defines the distance between adjacent windows in the $Spectrogram$ process, influencing time resolution and computational complexity. A smaller window hop increases overlap and time resolution but also raises computational complexity.

\subsection{Architecture description}
Our ASR model is a 2D convolutional neural network designed for spectrogram-based audio processing. It is inspired by the ResNet~\cite{resnet, audioresnet} architecture and consists of a series of normal and transition blocks. 
Our contribution centers around the challenges posed by low SNR and non-linear frequency response of the DPS. Our ASR model introduces \textit{learnable} \textbf{\textit{denoising autoencoder}} and \textbf{\textit{equalization layers}} to the {ResNet} architecture to provide a robust solution to the specific issues posed by the sensor.
The normal blocks contain a residual connection, while the transition blocks are responsible for reducing the spatial dimensions and increasing the number of channels. The model employs SubSpectralNorm~\cite{subspectral}, a normalization technique that applies batch normalization across sub-bands in the frequency domain, improving the model's performance on spectrogram-based tasks. ResNet also utilizes depth-wise separable convolutions for increased efficiency and reduced computational complexity. The final layers include a depth-wise convolution, a 1x1 convolution, and a head convolution for classification. The model is suitable for various audio tasks, such as speech recognition or sound event detection, where the input is a time-frequency representation of the audio signal.

\section{Performance on General Speech-Focused Sentences}
\label{appendix:performance}
Table~\ref{table:generalMSR} shows the average WER and MOS of each
ground truth sentence over the responses of the 18 volunteers.

\begin{table}[h!]
		\footnotesize
		\centering

\begin{tabular}{|p{6.3cm}|p{0.5cm}|p{0.5cm}|}
\hline
\graycell \textbf{Ground Truth Sentence }& \graycell \textbf{WER} &\graycell \textbf{MOS}\\ 
\hline
An object will remain at rest. & 0.54 & 3.89 \\ \hline
For every action in nature there is an equal and opposite reaction. & 0.49 & 3.89 \\ \hline
Here is my password  7 5 6 2 3. & 0.29 & 4.25 \\ \hline
He said the weather will be cold today. & 0.43 & 4.06 \\ \hline
The white egg is bigger than the green one. & 0.21 & 4.36 \\ \hline
A picture is worth a thousand words. & 0.20 & 4.22 \\ \hline
A journey of a thousand miles begins with a single step. & 0.31 & 4.11 \\ \hline
A bird in the hand is worth two in the bush. & 0.30 & 3.75 \\ \hline
Actions speak louder than words. & 0.27 & 4.50 \\ \hline
Never put off until tomorrow what you can do today. & 0.43 & 3.86 \\ \hline
\textbf{Average} & 0.35 & 4.09 \\ \hline
\multicolumn{3}{l}{${\textbf{NB:}}$  IRB exemption approval obtained to conduct the survey.} \\\end{tabular}
\caption{\label{table:generalMSR}Performance on general speech-focused sentences.}
			
\end{table}


\section{Semiconductor Manufacturing Process}
\label{appendix:semiconductorInfo}

\subsection{{Process description}}
\label{appendix:semiconductorProcessDesc}
\textit{As part of our survey, we provided volunteers with the following information to help them understand the manufacturing process.}
Semiconductor manufacturing is the process of creating electronic components from semiconductor materials, such as silicon. These components are used in a wide range of electronic devices, from smartphones to computers to cars. The key stages of this process are explained as follows~\cite{kamal2022}:

\begin{itemize}
    \item Wafer fabrication: In this step, silicon wafers are created by growing a single crystal of silicon and slicing it into thin, circular wafers.
    \item Photolitography: A beam of UV light of specific wavelength is passed through a template mask onto a layer of photoresistive material on the waffer to carve a pattern onto the material.
    \item Etching: Chemicals are used to remove material from the wafer, leaving only the desired pattern behind.
    \item Deposition: A layer of material is deposited onto the wafer, either by chemical vapor deposition or physical vapor deposition, to create specific features.
    \item Packaging: The individual chips are cut from the wafer and packaged into the final product.
\end{itemize}

\section{Future Work}
\label{appendix:futureWork}

\textbf{Effect of Distance on ASR Model Performance:} Based on the findings in Table~\ref{tab:factorsASR}, we have discovered that the distance between the DPS and the sound source affects the accuracy of our ASR model. 
With increasing distance, the sensor registers a weaker signal strength. Although this presents a limitation to the current effectiveness of the BaroVox approach, further study is warranted. As the STFT plots in Fig.~\ref{fig:distance} suggest, even with reduced signal strength due to increased distance (from 0.5m to 2m), the pressure sensor continues to detect speech signals. There's potential for refining the model by accounting for these distance variations.

\textbf{Vocabulary Limitations of the Current Model:} The current model is bound by vocabulary constraints, limiting its speech recognition capability. A strong attacker can enhance the capabilities by creating a complete speech-to-text translation system. Addressing challenges such as frequency spectrum unfolding for under-sampled pressure sensor data could provide a pathway to navigate the sampling rate constraints of the DPS~\cite{nader2011harmonic, nader2011unfolding}.

\textbf{Exploration of Advanced ASR Models:} We plan to build upon other advanced speech recognition models~\cite{chung2021w2vbert,zhang2022pushing, hasani2022liquid}. Leveraging these models alongside digital signal processing techniques may offer avenues to mitigate existing limitations.

\textbf{Impact of Sampling Rate}: In this study, the exploration of sampling rate's influence on model performance was limited due to space constraints. An in-depth investigation into this aspect remains a subject for future research.

\begin{figure}[h]
\centering
\includegraphics[width=0.495\textwidth,height=0.13\textheight]{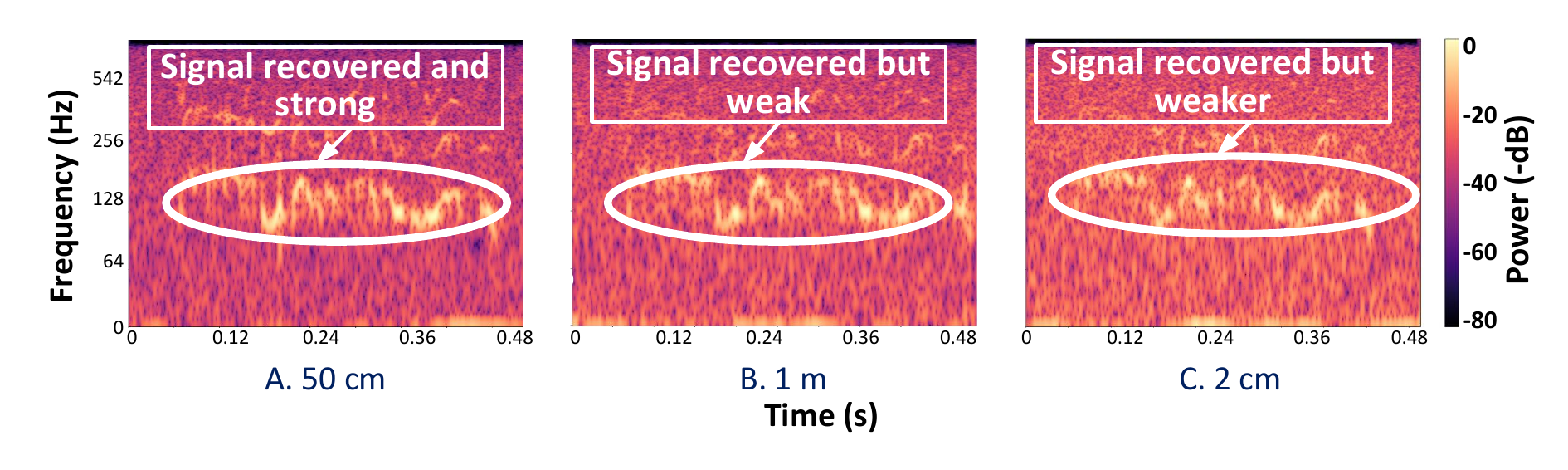}

\vspace{-1.000em}

\caption{STFT plot of speech signals recovered from pressure readings when the sound source is put at various distances from the sensor.
}
\label{fig:distance}

\end{figure}

\end{document}